\documentclass[journal]{IEEEtran}

\usepackage{amsmath,amssymb,bm}
\usepackage{array}
\usepackage{booktabs}
\usepackage{multirow}
\usepackage{cite}
\usepackage{url}
\usepackage{enumitem}
\newcommand{\Tx}{\mathrm{Tx}}
\newcommand{\Rx}{\mathrm{Rx}}
\newcommand{\TR}{\mathrm{TR}}
\newcommand{\RT}{\mathrm{RT}}
\newcommand{\UNK}{\mathrm{UNK}}
\newcommand{\NAV}{\mathrm{NAV}}
\newcommand{\TM}{\mathrm{TM}}
\newcommand{\TC}{\mathrm{TC}}
\newcommand{\Aset}{\mathcal{A}}
\newcommand{\Nset}{\mathcal{N}}
\newcommand{\Sset}{\mathcal{S}}
\newcommand{\Gset}{\mathcal{G}}
\newcommand{\Uset}{\mathcal{U}}

\newcommand{\Eset}{\mathcal{E}}
\newcommand{\Lset}{\mathcal{L}}
\newcommand{\Tset}{\mathcal{T}}

\usepackage{graphicx}
\usepackage{amsmath,amssymb}
\usepackage{booktabs}
\usepackage{array}
\usepackage[linesnumbered,ruled,vlined]{algorithm2e}
\usepackage{mathtools}

\begin{document}
	
	\title{Half-Slot Link Scheduling for Open GNSS with $x$-Tx $y$-Rx Multibeam Phased Array Antennas}
	
		\author{
				\IEEEauthorblockN{
						Huan Yan\IEEEauthorrefmark{1},
						Kanglian Zhao\IEEEauthorrefmark{1},
						Wenfeng Li\IEEEauthorrefmark{1},
						Jinjun Zheng\IEEEauthorrefmark{4},
						Ziyu Wang\IEEEauthorrefmark{4},
						Changgang Zheng\IEEEauthorrefmark{4},
					}
				
				\IEEEauthorblockA{
						\IEEEauthorrefmark{1} Nanjing University, Nanjing, China \\
						\IEEEauthorrefmark{4} China Academy of Space Technology, Beijing, China \\
					}
					\thanks{
						(Corresponding authors: Kanglian Zhao; Jinjun Zheng.)}
			}
	\maketitle
   \begin{abstract}
		Global Navigation Satellite Systems (GNSSs) are evolving from closed navigation constellations toward open space infrastructure capable of serving extensive spacecraft users. This evolution requires the constellation to support navigation ranging, telemetry (TM) return, and telecommand (TC) delivery for users while preserving its own ranging and communication functions.
		Existing phased-array payloads onboard BeiDou satellites provide one transmit (Tx) beam and one receive (Rx) beam. Such a configuration was designed mainly for intra-constellation operation, and can become a resource bottleneck when a large number of users share the same link opportunities. Upgrading GNSS satellites to phased-array terminals with $x$-Tx and $y$-Rx beams is therefore a natural enhancement. This paper develops a staged dynamic half-slot link scheduling framework (SDHSS) for open GNSS with $x$-Tx $y$-Rx multibeam phased-array antennas. The framework models navigation, TM, and TC as three different services, reserves the essential ranging and ground-access opportunities required by the GNSS constellation, and then allocates the remaining half-slot beam resources according to current service demand. To support multibeam concurrency, feasible links are selected through dynamic service-layer ordering and beam-copy matching. Simulations based on a BeiDou constellation show that the proposed method protects GNSS baseline operation and provides stronger external navigation, TM, and TC performance than fair contact planning. The results further show that additional Tx and Rx beams can be effectively converted into multi-service capability. This work establishes a link-scheduling approach for extending GNSS from constellation-internal operation to large-scale spacecraft service support.
   \end{abstract}
	
	\begin{IEEEkeywords}
		Global Navigation Satellite Systems (GNSSs), Multibeam phased array antenna, Link scheduling.
		
	\end{IEEEkeywords}

	\section{Introduction}
	
	Global Navigation Satellite Systems (GNSSs) provide positioning, navigation, and timing services to ground users and increasingly rely on inter-satellite links for ranging, time synchronization, and information exchange beyond direct ground-station (GS) coverage~\cite{1}. Precise orbit determination requires GNSS satellites to obtain ranging from many counterparts within short time windows, which favors rapidly reconfigurable phased-array (PA) antennas with electronic beam steering~\cite{15}. BeiDou satellites currently employ a Ka-band PA terminal with one transmit (Tx) beam and one receive (Rx) beam~\cite{2}. Because strong transmission can interfere with weak reception on the same terminal, Tx and Rx operate in time division manner~\cite{2}.	
	Prior GNSS link-scheduling studies all treat one slot, typically 3~s, as the basic unit for establishing a bidirectional phased-array link (PAL)~\cite{5,6,7,8,9,10,21}. Consider two nodes \(a\) and \(b\). During the first half of the slot, \(a\) transmits while \(b\) receives, forming the directed link \(a\rightarrow b\). During the second half, their roles are reversed and the directed link \(b\rightarrow a\) is established. These two opposite directed transmissions together form one bidirectional PAL in the slot. By changing PAL counterparts across consecutive slots, a GNSS satellite can poll different satellites and GSs for two-way ranging and information exchange.
	
	Current GNSS PALs are used mainly for constellation-internal functions. Although several studies have explored the potential of GNSS PALs to assist spacecrafts in space, they typically focus on one or only a few specific spacecraft~\cite{3,4,11,13}. 
	However, the rapid electronic steering of PA antennas suggests a broader role: GNSS could evolve from a closed navigation constellation into space infrastructure that systematically serves a large population of external spacecraft. We refer to this infrastructure-oriented operating mode as \emph{open GNSS}. In this paper, external spacecraft are classified as navigation users requiring ranging support, telemetry (TM) users returning data to the ground, and telecommand (TC) users receiving commands from the ground.
	
	Serving hundreds or even thousands of users can exceed the capability of conventional one-Tx one-Rx payloads. A natural enhancement is to equip GNSS satellites with \(x\)-Tx \(y\)-Rx multibeam PA terminals~\cite{19,20}, which can activate up to \(x\) Tx beams when transmitting or \(y\) Rx beams when receiving. This upgrade not only increases the number of simultaneous links, but also changes the link scheduling problem in three coupled ways.
	
	\paragraph{Scale surge and heterogeneous service demands}
	Open GNSS expands scheduling from a constellation of tens of navigation satellites to a network containing hundreds or potentially thousands of users. At the same time, PAL counterparts can change on the order of seconds, while multibeam concurrency allows each satellite to establish several contacts within the same scheduling instant. The resulting problem is therefore both large-scale and highly dynamic. More importantly, the users do not consume link resources in the same way: navigation requires two-way ranging opportunities, TM generates continuous traffic toward the ground, and TC requires low-delay reachability from the ground to potential targets. The scheduler must preserve GNSS-internal ranging and communication while allocating shared resources among these heterogeneous services. These characteristics make previous integer linear programming~\cite{5,6} and heuristic schemes~\cite{7,8,21} difficult to apply directly as a monolithic solution.
	
	\paragraph{Half-slot granularity and phase coupling}
	Conventional GNSS scheduling treats one full slot as the basic unit of a bidirectional contact~\cite{5,6,7,8,9,10,21}. This is suitable for two-way ranging, because the two opposite transmission directions jointly complete one ranging interaction. Communication, however, does not always require the two directions to be paired with the same counterpart. Scheduling at half-slot granularity allows the two halves of a slot to be assigned independently to different directed communication actions, thereby exposing transmission opportunities that slot-level bidirectional scheduling cannot use.
	This finer granularity also makes the Tx/Rx phase order an explicit scheduling decision. Under the conventional one-Tx one-Rx bidirectional setting, a satellite communicates with only one counterpart in a slot; therefore, whether the satellite transmits first or receives first has little effect on the resulting contact. With multibeam terminals and half-slot directed scheduling, this is no longer true. A satellite may receive from several upstream nodes in one half-slot and transmit to several downstream nodes in the other, so the receive-then-transmit and transmit-then-receive orders enable different sets of directed link combinations and different within-slot forwarding patterns. Because all concurrent links of a satellite must follow the same Tx/Rx mode in each half-slot, selecting half-slot links also constrains the phase order, while the phase order in turn determines which directed links are feasible. Half-slot link allocation and Tx/Rx phase ordering must therefore be scheduled jointly.
	\paragraph{Service-driven asymmetry topology}
	TM and TC require different directional structures. TM is continuously generated and may accumulate, so the network must provide sufficient throughput and low-delay paths toward the ground. TC is sporadic and small in volume, but requires low-delay reachability from GSs to remote targets. Conventional full-slot scheduling~\cite{5,6,7,8,9,10,21} produces structurally symmetric uplink and downlink contacts, even though the two services place different demands on direction, path formation, and scheduling value. Open-GNSS scheduling therefore needs service-aware directed links rather than a symmetric contact topology.
	
	These challenges lead to the problem studied in this paper: how to schedule the \(x\)-Tx \(y\)-Rx beam resources of GNSS PA terminals at half-slot granularity while preserving GNSS baseline functions and serving large-scale navigation, TM, and TC users. Accordingly, we make the following contributions:
	
	\begin{itemize}[leftmargin=*]
		\item \textit{Half-slot multibeam scheduling formulation.}
		We formulate open-GNSS link scheduling with half-slots and individual Tx/Rx beams as resource units. The model captures directed communication actions, Tx/Rx phase ordering, and multibeam concurrency within one scheduling framework.
		
		\item \textit{Heterogeneous service models.}
		Navigation, TM, and TC are modeled according to their operational requirements rather than a common contact metric. Navigation is characterized by distinct ranging opportunities and phase constraints for long-range users; TM is modeled as periodic traffic relayed toward GSs; and TC is represented by an event-replica process that measures low-delay reachability from the ground to potential targets.
		
		\item \textit{Staged dynamic scheduling with beam-copy matching.}
		We propose a two-stage framework that first reserves a sparse contact template for GNSS baseline operation and then allocates residual half-slot resources according to navigation deficits, TM backlogs, and TC event states. Dynamic service-layer ordering adapts resource competition to current demand, while beam-copy maximum-weight matching exploits multibeam concurrency without violating Tx/Rx resource conflicts.
		
		\item \textit{BeiDou-based evaluation against conventional contact planning.}
		We compare the proposed scheduler with fair contact planning~\cite{12} under consistent service metrics. Simulations show that the proposed method preserves GNSS baseline performance, substantially improves TM and TC delivery, and converts additional Tx/Rx beams into scalable navigation and communication services for external users.
	\end{itemize}

\section{Network Model and Scheduling Problem}
\label{sec:network_problem}

\subsection{Network Entities}
\label{subsec:network_entities_roles}

The open GNSS network consists of GNSS satellites, ground stations (GSs), and external users. Let \(\mathcal{S}\) and \(\mathcal{G}\) denote the sets of GNSS satellites and GSs, respectively. According to their requested services, users are divided into navigation users \(\mathcal{U}^{\mathrm{N}}\), telemetry (TM) users \(\mathcal{U}^{\mathrm{M}}\), and telecommand (TC) users \(\mathcal{U}^{\mathrm{C}}\). The complete node set is
\[
\mathcal{V}
=
\mathcal{S}\cup\mathcal{G}
\cup\mathcal{U}^{\mathrm{N}}
\cup\mathcal{U}^{\mathrm{M}}
\cup\mathcal{U}^{\mathrm{C}}.
\]

GNSS satellites form the infrastructure core of the network. They support constellation-internal ranging, time synchronization, and information exchange, while also acting as communication nodes that generate TM and may receive TC instructions from the ground segment. In the open-GNSS setting, they further provide ranging access and traffic-relay capability for external users.

GSs provide the interface between the space segment and the ground control segment. They collect TM returned by GNSS satellites and TM users, and originate TC instructions destined for GNSS satellites or TC users.

The three user classes interact with the GNSS infrastructure differently. Navigation users establish two-way ranging interactions with GNSS satellites; TM users generate data that must be returned to GSs through the GNSS; and TC users are destinations of instructions originating from the ground segment. These distinct service roles determine the link directions and state evolution modeled in Sec.~\ref{subsec:service_modeling}.

\subsection{FSA-Based Visibility and Anchor Satellite Partition}
\label{subsec:fsa_visibility_anchor}

\begin{figure}[t]
	\centering
	\includegraphics[width=0.95\linewidth]{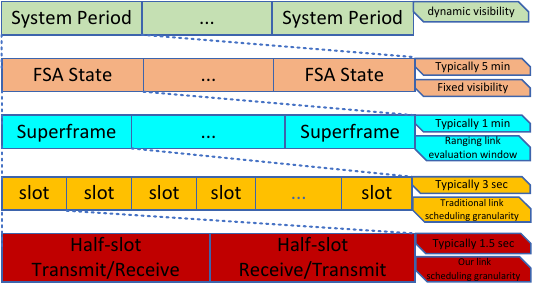}
	\caption{FSA-based visibility and half-slot scheduling granularity.}
	\label{fig:topology_model}
\end{figure}

Two nodes are considered visible if their line of sight is unobstructed and each lies within the antenna coverage of the other.
Visibility is a prerequisite for establishing a link.
Since inter-node visibility varies continuously with node motion, the operational period is partitioned into equal-duration finite-state-automaton (FSA) states~\cite{17}, as shown in Fig.~\ref{fig:topology_model}. A node pair is treated as visible in FSA state \(f\) only if it remains continuously visible throughout that interval; otherwise, it is treated as invisible for the entire interval. This conservative rule converts the time-varying geometry into a static visibility graph for each FSA state. Define
\begin{equation}
	\chi_{ij}(f)\in\{0,1\},\quad i,j\in\mathcal{V},
	\label{eq:unified_visibility}
\end{equation}
where \(\chi_{ij}(f)=1\) denotes continuous visibility between nodes \(i\) and \(j\) throughout FSA state \(f\).

In each FSA state, a satellite visible to at least one GS is classified as an anchor satellite:
\begin{equation}
	\mathcal{A}_f
	=
	\{s_i\in\mathcal{S}\mid
	\exists g_j\in\mathcal{G},\ \chi_{s_i g_j}(f)=1\}.
	\label{eq:anchor_set}
\end{equation}
The remaining satellites form the non-anchor set,
\begin{equation}
	\mathcal{N}_f=\mathcal{S}\setminus\mathcal{A}_f.
	\label{eq:nonanchor_set}
\end{equation}
Because visibility changes with \(f\), the same satellite may change roles across FSA states. Only anchor satellites have direct GS access. Consequently, TM from non-anchor satellites or TM users admitted by them must generally reach GSs through anchor satellites, whereas TC enters the constellation through anchors before further dissemination.

\subsection{Superframe--Slot--Half-Slot Multibeam Model}
\label{subsec:superframe_slot_halfslot_model}

Within each FSA state, scheduling uses the nested superframe--slot--half-slot structure shown in Fig.~\ref{fig:topology_model}. A superframe is the horizon for evaluating ranging diversity: repeated measurements with the same counterpart within one superframe count as one effective ranging opportunity. Each superframe contains \(T\) slots, and each slot contains two half-slots. A slot retains the conventional bidirectional PAL structure required by two-way ranging, whereas a half-slot is the basic unit for scheduling a directed link.

Let \(\mathcal{T}\) denote the set of slot indices in one superframe, and let \(\mathcal{H}\) denote the corresponding set of half-slots, where each half-slot is identified by its slot index \(t\) and half-slot index \(h\). Specifically,
\begin{equation}
	\begin{aligned}
		\mathcal{T}&=\{0,1,\ldots,T-1\},\\
		\mathcal{H}&=\{(t,h)\mid t\in\mathcal{T},\ h\in\{0,1\}\},
		\qquad \tau=2t+h .
	\end{aligned}
	\label{eq:time_sets}
\end{equation}
Here, \(\tau\) is the ordered half-slot index, which maps the two-dimensional pair \((t,h)\) to a single chronological index.

Each node \(v\) has \(X_v\) Tx beams and \(Y_v\) Rx beams; users use \(X_u=Y_u=1\) by default. A GNSS satellite cannot transmit and receive simultaneously in the same half-slot. Let
\[
\theta_s(t,h)\in\{0,1\},\qquad s\in\mathcal{S},
\]
denote the phase of satellite \(s\), where \(1\) and \(0\) represent Tx and Rx modes, respectively. The two half-slots of a slot use opposite modes:
\begin{equation}
	\theta_s(t,1)=1-\theta_s(t,0),
	\quad \forall s\in\mathcal{S},~t\in\mathcal{T}.
	\label{eq:phase_consistency}
\end{equation}
Thus, \(\theta_s(t,0)=1\) gives the transmit-then-receive order \(\TR\), while \(\theta_s(t,0)=0\) gives the receive-then-transmit order \(\RT\).

User terminals obey the same half-slot Tx/Rx time-division restriction and, under the default single-beam setting, communicate with at most one counterpart at a time. Considering the stronger processing capability of GS terminals, their Tx and Rx resources are modeled as independent and may operate simultaneously, subject to their respective beam number limits.

In this paper, a scheduled action is a directed link \(i\rightarrow j\) in one half-slot and consumes one Tx beam at node \(i\) and one Rx beam at node \(j\). The nested model preserves slot-level two-way ranging while enabling concurrent half-slot directed communication with multibeam resources.

\subsection{Service Models for Navigation, Telemetry, and Telecommand}
\label{subsec:service_modeling}

Navigation, telemetry (TM), and telecommand (TC) impose fundamentally different requirements on link scheduling. Navigation values two-way ranging with geometrically diverse satellites, TM requires sustained data delivery toward the ground, and TC requires low-delay reachability from the ground to potentially remote targets. We therefore model the three services separately so that their distinct states and scheduling values can be preserved.

\paragraph{Navigation service}
For a navigation user, service quality is measured by the number of distinct GNSS satellites with which valid ranging interactions are completed within one superframe. Navigation users are divided into short-range and long-range classes, denoted by \(\mathcal{U}^{\mathrm{NS}}\) and \(\mathcal{U}^{\mathrm{NL}}\), according to their distance from the GNSS constellation.

For a short-range user, the propagation delay is negligible relative to the slot duration. A valid service therefore requires two opposite directed half-slot links between the user and one GNSS satellite within the same slot. Either \(\TR\) or \(\RT\) can complete this two-way interaction.

\begin{figure}[t]
	\centering
	\includegraphics[width=0.95\linewidth]{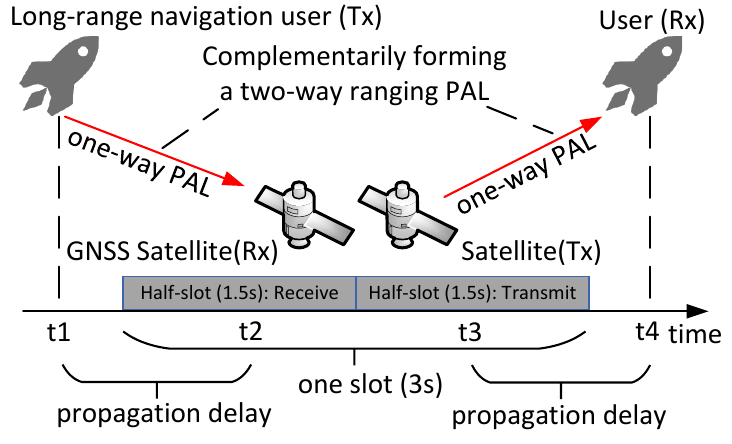}
	\caption{Satellites use fixed \(\RT\) phase to serve long-range navigation users.}
	\label{fig:SatRTMode}
\end{figure}

For a long-range user, such as a cislunar spacecraft, the propagation delay may be comparable to the slot duration. As illustrated in Fig.~\ref{fig:SatRTMode}, an \(\RT\) satellite can receive the user's ranging signal in the first half of the slot and transmit the response in the second half, allowing the complete user--satellite--user interaction to finish within one slot after accounting for propagation delay. Under \(\TR\), the response transmission must wait for a later slot, causing the ranging process to span multiple slots. Prior work that does not explicitly schedule the satellite transceiver phase therefore adopts conservative multi-slot contacts for long-range navigation~\cite{3}. To avoid this additional slot occupation, long-range navigation service in this paper requires the serving satellite to use the \(\RT\) phase order.

\paragraph{Telemetry service}
TM is modeled as real, volume-conserving traffic that is periodically generated and must eventually reach a GS. Both GNSS satellites and TM users may generate TM data. A TM user may upload data to any visible GNSS satellite. Data held by an anchor satellite can be downlinked directly to a GS, whereas data held by a non-anchor satellite must first be forwarded to an anchor satellite and then downlinked. GNSS-satellite TM follows the same groundward structure. Because TM is high-volume traffic, this model excludes anchor--anchor and non-anchor--non-anchor TM forwarding, avoiding additional hops that consume beam resources and link capacity without advancing the prescribed ground-access structure. The scheduler must therefore decide which TM data to admit, forward, and downlink so as to improve delivery completion and reduce end-to-end delay.

\paragraph{Telecommand service}
TC instructions are event-driven, unpredictable, and delay-sensitive. A real TC may be issued from the ground at an arbitrary time and must reach a target GNSS satellite or TC user quickly. The scheduling objective is therefore not sustained TC throughput, but the continuous maintenance of low-delay reachability from the ground segment to potential targets.

To make this latent reachability observable, TC is modeled as a virtual event-replica process. Small virtual TC events are generated periodically for potential targets and initially made available at the GSs. These events are probes rather than real command payloads. When a node holding an unfinished event establishes a useful directed link to another node, the event can be copied to the receiver while the sender retains its copy. The holder set therefore expands over time, allowing different replicas of the same event to explore multiple propagation paths. This replication model is appropriate for TC because command traffic is small and sporadic compared with TM, so reachability can be emphasized without modeling TC as a bulk flow.

A virtual event enters the constellation through a GS--anchor link and may then propagate over scheduled inter-satellite links and satellite-TC user links. Replicas may follow different paths until one reaches the target GNSS satellite or a replica-holding satellite delivers the event to its target TC user. The first replica arrival marks the event as completed. Its completion delay represents the low-delay path that the current schedule could provide to a real TC issued at the corresponding time. In this way, unpredictable TC arrivals are transformed into observable reachability-maintenance tasks while preserving the first-arrival semantics of command delivery.

\paragraph{Half-slot transmission completion model}
Considering that the complete transmission of data in reality requires a certain amount of time
, TM and TC state evolution follows a half-slot transmission-completion rule. If node \(a\) transmits information to node \(b\) in half-slot \(\tau\), that information becomes available at \(b\) only from half-slot \(\tau+1\). Hence, TM received in \(\tau\) cannot be forwarded again within the same half-slot, and a TC replica created in \(\tau\) cannot be further replicated or delivered until a later half-slot. TM backlog propagation and TC holder-set evolution therefore proceed step by step across half-slot boundaries rather than through instantaneous same-half-slot cascades. For navigation, a serving satellite is added to the distinct-ranging record only after the required two-way ranging interaction is completed.

\subsection{Ideal Scheduling Problem}
\label{subsec:ideal_scheduling_problem}

Under a fixed FSA visibility, the scheduler plans one superframe of half-slot directed links. Given the visibility, node beam resources, and navigation, TM, and TC demands, the decision consists of the directed-link schedule \(\Pi\) over all half-slots and the corresponding Tx/Rx phase arrangement \(\theta\). For each \((t,h)\in\mathcal{H}\), selected links must satisfy visibility, endpoint phase compatibility, and Tx/Rx beam limits, while their service effects follow the navigation, TM, TC, and half-slot completion models defined above. Since visibility is fixed within one FSA state, the resulting superframe schedule can be reused by the remaining superframes in that FSA.

The schedule must jointly protect GNSS baseline operation and serve heterogeneous users. GNSS performance depends on inter-satellite ranging diversity and basic TM/TC access to the ground; navigation users require sufficient distinct ranging counterparts; TM users require high delivery completion and low end-to-end delay; and TC users require low-delay reachability from the ground to its targets. We summarize these coupled objectives by
\begin{equation}
	\mathbf{P_0}:\quad
	\max_{\Pi,\theta}\; \mathcal{J}(\Pi(\theta)),
	\label{eq:ideal_scheduling_objective}
\end{equation}
where \(\mathcal{J}(\Pi(\theta))\) jointly characterizes GNSS baseline performance, navigation ranging diversity, TM delivery efficiency, and TC reachability delay.

Directly solving \(\mathbf{P_0}\) is difficult for two reasons. First, all services compete for the same half-slot multibeam resources but have different state variables, value structures, and temporal evolution, making a unified monolithic formulation highly coupled. Second, the problem scale is large: hundreds of nodes generate a large candidate set, each half-slot requires selecting a conflict-free set of directed links rather than a single contact, and navigation accumulation, TM backlog propagation, and TC replica diffusion couple decisions across successive half-slots. Joint optimization over all services and half-slots is therefore computationally prohibitive. These two difficulties motivate the staged and state-driven scheduling design developed in the next section.

\section{Staged Dynamic Half-Slot Scheduling Design}
\label{sec:design}

\subsection{Design Overview}
\label{subsec:algorithm_interface}

To address the coupling and scale of \(\mathbf{P_0}\), we propose a staged dynamic half-slot scheduling framework, termed \emph{SDHSS}. Its core principle is to protect GNSS baseline operation first and then allocate the remaining multibeam resources according to the current service state.

Stage~1 constructs a sparse baseline contact template \(\Pi^{\mathrm{base}}\) over GNSS satellites and GSs. The template reserves the link opportunities required for inter-satellite ranging diversity and the minimum contact opportunities needed to maintain GNSS-satellite TM return and TC reachability, preventing large-scale service competition from degrading essential GNSS operation. Stage~1 plans slot-level bidirectional contacts; each selected contact is subsequently oriented into two opposite half-slot actions, which determine the corresponding satellite phases and occupy the associated Tx/Rx beam resources. Stage~2 then performs dynamic scheduling only over the phase-feasible residual resources left by these oriented Stage~1 actions.

Stage~2 performs traffic-aware scheduling over the residual resources of each half-slot. At the beginning of half-slot \(\tau\), the scheduler observes the current navigation deficits, TM backlogs, and unfinished TC event states. It then orders service layers by urgency and, layer by layer, constructs feasible directed-link candidates and selects non-conflicting actions through beam-copy matching. Selected actions immediately reserve phase and beam resources, while their service effects are committed at the half-slot boundary according to the transmission-completion model. This state-driven procedure replaces a monolithic optimization over all services and half-slots with a sequence of local scheduling decisions.

The two stages separate scheduling responsibilities rather than traffic types. Oriented Stage~1 actions may carry user TM data or TC event replicas whenever their directions and endpoints provide useful service transitions. Conversely, Stage~2 actions may serve GNSS-satellite TM and TC as well as external users. Thus, the service contribution of a scheduled link is determined by its direction, endpoints, capacity, and current service state, rather than by the stage that created it.

Algorithm~\ref{alg:staged_half_slot_scheduler} summarizes the overall procedure for one scheduling superframe. The following subsections detail Stage~1 baseline reservation and orientation, Stage~2 service-layer ordering, beam-copy matching, and reservation--commit semantics. Exact runtime calculations for service states, urgency, marginal weights, planned items, and delay recording are provided in the supplementary material.

\begin{algorithm}[t]
	\caption{Staged Dynamic Half-Slot Scheduling}
	\label{alg:staged_half_slot_scheduler}
	\KwIn{FSA state \(f\); visibility \(\chi(f)\); node sets and anchor partition \((\Aset_f,\Nset_f)\); beam resources \(X_v,Y_v\); navigation, TM, and TC demands.}
	\KwOut{Half-slot directed-link schedule \(\Pi(\theta)\).}
	
	Initialize \(\Pi(\theta)\leftarrow\emptyset\) and the service states\;
	
	\BlankLine
	\tcp{\footnotesize Stage 1: baseline GNSS reservation}
	Construct the slot-level contact template
	\(\Pi^{\mathrm{base}}\) for GNSS baseline operation\;
	
	\BlankLine
	\tcp{\footnotesize Stage 2: dynamic residual resource schedule}
	\ForEach{\((t,h)\in\Tset\times\{0,1\}\)}{
		Set \(\tau\leftarrow 2t+h\)\;
		
		Observe the current service-state snapshot \(\Omega_\tau\)\;
		
		\If{\(h=0\)}{
			Determine the orientations of the Stage~1 bidirectional contacts
		    according to the Stage~1
			orientation rules and the slot-start state
			\(\Omega_{2t}\), thereby fixing the opposite directed
			actions in \((t,0)\) and \((t,1)\)\;
		}
		
		Apply Stage~1 one-way actions for half-slot \((t,h)\):
		reserve their phase/beam resources and reserve the eligible
		TM/TC items planned for transmission\;
		
		Obtain the residual phase-feasible Tx/Rx resources available
		to Stage~2\;
		
		Compute service-layer urgencies and determine the execution order\;
		
		\ForEach{service layer \(\ell\) in the ordered sequence}{
			Generate feasible candidates, select actions by beam-copy
			matching, and add them to \(\Pi\)\;
			
			Update the resource ledger and reserve the planned
			service effects\;
		}
		
		Commit TM/TC state updates and completed ranging records
		at the half-slot boundary\;
	}
	
	\Return \(\Pi(\theta)\);
\end{algorithm}

\subsection{Stage~1: Baseline GNSS Resource Reservation}
\label{subsec:stage1_baseline_reservation}

Stage~1 reserves a sparse set of constellation-only contacts before large-scale services compete for resources. Its purpose is to protect two baseline capabilities: sufficient inter-satellite ranging diversity and recurring contact opportunities for GNSS-satellite TM return and TC reachability. Because ranging is inherently bidirectional, Stage~1 uses slot-level bidirectional contacts as its decision unit. The resulting template \(\Pi^{\mathrm{base}}\) contains only satellite--satellite and satellite--GS contacts; user nodes do not enter the Stage~1 optimization. After optimization, each selected contact is oriented into two opposite half-slot actions and occupies the corresponding phase and beam resources.

For FSA state \(f\), define the visible baseline candidate-edge set as
\begin{equation}
	\Eset^{\mathrm{B}}_f=
	\left\{
	\{i,j\}\,\middle|\,
	\substack{
		i,j\in\Sset\cup\Gset,\ i\ne j,\\
		\chi_{ij}(f)=1,\ \{i,j\}\not\subseteq\Gset
	}
	\right\},
	\label{eq:stage1_candidate_edge_set}
\end{equation}
which contains visible satellite--satellite and satellite--GS edges. Let \(\Eset_f^{\mathrm{SS}}\subseteq\Eset_f^{\mathrm{B}}\) denote the satellite--satellite subset. For \(e\in\Eset_f^{\mathrm{B}}\), let \(x_e(t)\in\{0,1\}\) indicate whether edge \(e\) is selected in slot \(t\). For \(e\in\Eset_f^{\mathrm{SS}}\), let \(z_e\in\{0,1\}\) indicate whether the satellite pair is linked at least once in the superframe. Visibility is enforced by defining these variables only on \(\Eset_f^{\mathrm{B}}\).

To keep the baseline template sparse and make its half-slot orientation unambiguous, each satellite participates in at most one Stage~1 contact per slot. A GS may support up to
\(B_g^{\mathrm{pair}}=\min\{X_g,Y_g\}\)
simultaneous bidirectional baseline contacts. Thus,
\begin{align}
	\sum_{e\in\Eset^{\mathrm{B}}_f:\,s\in e}x_e(t)
	&\le 1,
	&&\forall s\in\Sset,\ t\in\Tset,
	\label{eq:stage1_satellite_degree}\\
	\sum_{e\in\Eset^{\mathrm{B}}_f:\,g\in e}x_e(t)
	&\le B_g^{\mathrm{pair}},
	&&\forall g\in\Gset,\ t\in\Tset.
	\label{eq:stage1_gs_degree}
\end{align}

The pair indicator \(z_e\) is linked to the slot-level decisions by
\begin{equation}
	z_e
	\le
	\sum_{t\in\Tset}x_e(t)
	\le
	Tz_e,
	\quad
	\forall e\in\Eset_f^{\mathrm{SS}}.
	\label{eq:stage1_ranging_link_indicator}
\end{equation}
Hence, repeated links with the same satellite pair count as one ranging counterpart within the superframe. Stage~1 requires
\begin{equation}
	\sum_{e\in\Eset_f^{\mathrm{SS}}:\,s\in e}z_e
	\ge L_{\min},
	\quad
	\forall s\in\Sset,
	\label{eq:stage1_lmin}
\end{equation}
where \(L_{\min}\) is the minimum number of distinct ranging counterparts required for each satellite.

The second protection objective is recurring access for GNSS-satellite TM and TC. Anchor--GS and non-anchor--anchor contacts can be written using a common next-hop access set:
\begin{equation}
	\mathcal{R}_p(f)=
	\begin{cases}
		\Gset, & p\in\Aset_f,\\
		\Aset_f, & p\in\Nset_f.
	\end{cases}
	\label{eq:stage1_relay_set}
\end{equation}
Accordingly, every anchor must contact a GS and every non-anchor must contact an anchor at least once within each sliding window of \(T_m\) slots:
\begin{equation}
	\sum_{t=\omega}^{\omega+T_m-1}
	\sum_{\substack{
			q\in\mathcal{R}_p(f):\\
			\{p,q\}\in\Eset_f^{\mathrm{B}}
	}}
	x_{\{p,q\}}(t)
	\ge 1,
	\quad
	\substack{
		\forall p\in\Aset_f\cup\Nset_f,\\
		\omega=0,\ldots,T-T_m
	}.
	\label{eq:stage1_window_connectivity}
\end{equation}
This bound limits prolonged loss of ground-contact opportunities for anchors and anchor-access opportunities for non-anchors.

As every Stage~1 contact reduces the resources available to dynamic service scheduling, the baseline template should satisfy the above protection requirements with as few reserved contacts as possible. We therefore solve
\begin{equation}
	\min_{x,z}\;
	\lambda_x
	\sum_{t\in\Tset}
	\sum_{e\in\Eset_f^{\mathrm{B}}}x_e(t)
	-
	\lambda_z
	\sum_{e\in\Eset_f^{\mathrm{SS}}}z_e,
	\label{eq:stage1_objective}
\end{equation}
where \(\lambda_x>0\), \(\lambda_z>0\), and \(\lambda_z\ll\lambda_x\). The dominant term minimizes baseline resource occupation, while the secondary term favors additional distinct satellite pairs among solutions with comparable contact cost. Since Stage~1 contains only constellation and GS nodes, this ILP remains independent of the number of external users.

The optimized slot-level contacts are then oriented into half-slot actions. If \(\{i,j\}\) is selected in slot \(t\), one direction is scheduled in \((t,0)\) and the reverse direction in \((t,1)\). For anchor-related contacts, the orientation follows the relay role of the anchor: an anchor--GS contact uses \(g\rightarrow a\) in \((t,0)\) and \(a\rightarrow g\) in \((t,1)\), while a non-anchor--anchor contact uses \(n\rightarrow a\) followed by \(a\rightarrow n\). Thus, the anchor adopts the \(\RT\) order, which is consistent with receiving upstream TM or TC information before using later transmission opportunities for groundward aggregation or outward dissemination.

For anchor--anchor and non-anchor--non-anchor contacts, neither endpoint has an inherent first-half direction. Their orientation is therefore chosen by comparing the useful TC-replica propagation gain of the two directions at the beginning of slot \(t\); the direction with the larger gain is scheduled in \((t,0)\), and the reverse direction in \((t,1)\). Ties are broken deterministically by letting the node with the smaller index transmit first. The exact direction-score calculation is given in the supplementary material.

After orientation, Stage~1 actions fix the corresponding satellite phases and reserve the associated Tx/Rx beams. They may also carry eligible TM data or TC event replicas according to the current service state. Stage~2 subsequently schedules only over the phase-feasible residual resources left by these actions.

\subsection{Stage~2: Service-Layer Decomposition and Dynamic Ordering}
\label{subsec:stage2_layer_decomposition}

Stage~2 allocates the phase-feasible residual resources left by Stage~1 according to the current navigation, TM, and TC states. Because these services differ in endpoints, state evolution, and scheduling value, their candidate actions are not mixed into one global pool. Instead, Stage~2 decomposes them into service layers, each representing one physically meaningful service transition:
\begin{equation}
	\begin{split}
		\Lset^{\mathrm{dyn}}=\{&
		\ell_{\NAV}^{\mathrm{NS}},~
		\ell_{\NAV}^{\mathrm{NL}},~
		\ell_{\TM}^{\mathrm{U2S}},~
		\ell_{\TM}^{\mathrm{N2A}},~
		\ell_{\TM}^{\mathrm{A2G}},\\
		&
		\ell_{\TC}^{\mathrm{G2A}},~
		\ell_{\TC}^{\mathrm{A2N}},~
		\ell_{\TC}^{\mathrm{D}}
		\}.
	\end{split}
	\label{eq:stage2_dynamic_layer_set}
\end{equation}
The two navigation layers serve short- and long-range users, respectively. The three TM layers represent user access, non-anchor-to-anchor transfer, and anchor-to-ground downlink. The three TC layers represent ground-to-anchor injection, anchor-to-non-anchor dissemination, and satellite-to-user delivery. A satellite-targeted TC event is completed when a feasible injection or dissemination action reaches its target satellite, whereas a user-targeted event requires \(\ell_{\TC}^{\mathrm{D}}\). Table~\ref{tab:stage2_candidate_rules} summarizes the direction and marginal-value basis of each layer.

\begin{table*}[t]
	\centering
	\caption{Stage~2 service-layer semantics and marginal-value sources.}
	\label{tab:stage2_candidate_rules}
	\footnotesize
	\begin{tabular}{p{0.12\textwidth}p{0.31\textwidth}p{0.49\textwidth}}
		\toprule
		Layer & Service transition & Positive marginal-value basis \\
		\midrule
		\(\ell_{\NAV}^{\mathrm{NS}}\) &
		short-range navigation user \(\leftrightarrow\) GNSS satellite &
		new distinct serving satellite; remaining ranging deficit and user priority \\
		\(\ell_{\NAV}^{\mathrm{NL}}\) &
		long-range navigation user \(\leftrightarrow\) \(\RT\)-compatible GNSS satellite &
		new distinct long-range ranging opportunity; remaining deficit and user priority \\
		\(\ell_{\TM}^{\mathrm{U2S}}\) &
		TM user \(\rightarrow\) GNSS satellite &
		positive user-side backlog; backlog amount and user priority \\
		\(\ell_{\TM}^{\mathrm{N2A}}\) &
		non-anchor \(\rightarrow\) anchor satellite &
		eligible source-side TM; amount, source priority, and data age \\
		\(\ell_{\TM}^{\mathrm{A2G}}\) &
		anchor satellite \(\rightarrow\) GS &
		eligible source-side TM; amount, source priority, and data age \\
		\(\ell_{\TC}^{\mathrm{G2A}}\) &
		GS \(\rightarrow\) anchor satellite &
		useful new holder or satellite-target completion; event priority, age, and replica/completion gain \\
		\(\ell_{\TC}^{\mathrm{A2N}}\) &
		anchor \(\rightarrow\) non-anchor satellite &
		useful frontier expansion or satellite-target completion; event priority, age, and replica/completion gain \\
		\(\ell_{\TC}^{\mathrm{D}}\) &
		satellite \(\rightarrow\) TC user &
		holder carries an unfinished event for the visible target user; request priority, age, and completion gain \\
		\bottomrule
	\end{tabular}
\end{table*}

The layers are not executed in a fixed priority order. Their relative importance changes with the runtime state: navigation deficits shrink as distinct measurements accumulate, TM pressure moves between users, non-anchors, and anchors, and an unfinished TC event may suddenly obtain a high-value propagation or completion opportunity. Therefore, Stage~2 recomputes layer urgency at the beginning of every half-slot.

Let \(\Omega_\tau\) denote the service-state snapshot available at the beginning of half-slot \(\tau\), including unsatisfied navigation demand, TM backlogs, and the holder states of unfinished TC events. For each \(\ell\in\Lset^{\mathrm{dyn}}\),
\begin{equation}
	u_\ell(\tau)=F_\ell(\Omega_\tau)\ge 0,
	\label{eq:stage2_layer_urgency}
\end{equation}
where \(F_\ell(\cdot)\) is the state-value function of layer \(\ell\). Sorting the layers by decreasing urgency gives
\begin{equation}
	\mathcal{O}_\tau=(\ell_1,\ell_2,\ldots,\ell_K),
	\quad
	u_{\ell_1}(\tau)\ge u_{\ell_2}(\tau)\ge\cdots\ge u_{\ell_K}(\tau),
	\label{eq:stage2_dynamic_order}
\end{equation}
where \(K=|\Lset^{\mathrm{dyn}}|\).
The urgency sources follow the service semantics. Navigation urgency is driven by the remaining distinct-ranging deficit. TM urgency is driven by backlog pressure and waiting time at the corresponding source side. TC urgency evaluates unfinished events at the event level using event age, destination priority, current holders, and feasible propagation or completion opportunities. Counting TC urgency by event rather than by replica multiplicity prevents heavily replicated events from being overvalued merely because they already have many copies. The exact urgency functions are given in the supplementary material.

The resulting order determines which service transitions obtain earlier access to residual resources in the current half-slot. Concrete links within each layer are then selected by the resource-controlled beam-copy matching procedure described next.

\subsection{Stage~2: Dynamic Layer Scheduling and Beam-Copy Matching}
\label{subsec:dynamic_layer_scheduling_halfslot}
\label{subsec:candidate_matching}

For half-slot \(\tau\), let \(R_v^{\Tx}(\tau)\) and \(R_v^{\Rx}(\tau)\) denote the Tx and Rx beams remaining at node \(v\) after Stage~1 reservation and previously executed Stage~2 layers. The physical capacities are \(C_v^{\Tx}=X_v\) and \(C_v^{\Rx}=Y_v\). Once an action is selected, its phase and beam resources are reserved immediately, whereas information availability still follows the half-slot transmission-completion model defined in Sec.~\ref{subsec:service_modeling}.

Dynamic ordering gives urgent layers earlier access to resources, but an early layer should not necessarily consume all beams of a multibeam node. Stage~2 therefore combines a rank-dependent fraction \(\alpha_r\in[0,1]\) with a layer-dependent fraction \(\beta_\ell\in[0,1]\). For layer \(\ell_r\) executed at rank \(r\),
\begin{equation}
	\eta_{\ell_r,r}
	=
	\min\{\alpha_r,\beta_{\ell_r}\}.
	\label{eq:stage2_effective_fraction}
\end{equation}
For direction \(m\in\{\Tx,\Rx\}\), the maximum number of beams that layer \(\ell_r\) may use at node \(v\) is
\begin{equation}
	\widehat{R}_{v,r}^{m}(\tau)=
	\begin{cases}
		\min\!\left\{
		R_v^{m}(\tau),
		\max\!\left\{
		1,\left\lfloor
		\eta_{\ell_r,r} C_v^{m}
		\right\rfloor
		\right\}
		\right\},
		&
		\substack{
			R_v^{m}(\tau)>0,\\
			\eta_{\ell_r,r}>0,
		}
		\\[1mm]
		0, & \text{otherwise}.
	\end{cases}
	\label{eq:stage2_layer_resource_cap}
\end{equation}
Thus, \(\widehat{R}_{v,r}^{m}(\tau)\) is a per-layer beam-use cap rather than a modification of the physical resource ledger. The real residual resource \(R_v^m(\tau)\) is deducted only after actions are selected. The one-beam minimum prevents a positive-fraction layer from being disabled by flooring on a small-beam node.

For an executed layer \(\ell\), let \(\mathcal{C}_\ell(\tau)\) denote its candidate action set. Each \(a\in\mathcal{C}_\ell(\tau)\) is a possible directed action with transmitter \(i_a\), receiver \(j_a\), capacity \(c_a\), and marginal weight \(w_a(\tau)\). A candidate is retained only if it satisfies the layer endpoint/direction rule in Table~\ref{tab:stage2_candidate_rules}, FSA visibility, current phase compatibility, residual-resource and layer-cap limits, and positive marginal contribution \(w_a(\tau)>0\). Exact filtering order and capacity-sliced planned-item selection are provided in the supplementary material.

The marginal weight measures the local service value of activating a candidate in the current half-slot. Navigation weights reward reduction of distinct-ranging deficits. TM user-access weights reflect user-side backlog, whereas satellite-side TM transfer weights reflect transferable amount, source priority, and batch age. TC weights reflect event priority and age together with the additional reachability gained by creating a useful holder or directly completing an event. Thus, TC value is based on event-level marginal reachability rather than replica count.

After candidate construction and weighting, Stage~2 uses maximum-weight matching~\cite{23} to select a conflict-free action set within the current layer. In a matching graph, each candidate action is represented by a weighted edge, and a feasible matching contains edges that do not share graph endpoints. Maximum-weight matching selects the feasible matching with the largest total edge weight. If physical nodes were used directly as graph endpoints, however, one node could be incident to at most one selected edge in a matching round, which would reduce an \(x\)-Tx \(y\)-Rx multibeam terminal to single-link usage.
To preserve multibeam concurrency, Stage~2 therefore expands the beam resources permitted by \(\widehat{R}_{v,r}^{m}(\tau)\) into individual resource copies. For \(m\in\{\Tx,\Rx\}\),
\begin{equation}
	\mathcal{B}_{v,r}^{m}(\tau)=
	\left\{
	(v,m,k)\mid
	k=1,\ldots,\widehat{R}_{v,r}^{m}(\tau)
	\right\}.
	\label{eq:beam_copy_set}
\end{equation}
Each copy represents one beam resource available to the current layer and can be incident to at most one selected matching edge. Since navigation consumes one Tx and one Rx beam of the same satellite for a two-way ranging service, its paired copy set is
\begin{equation}
	\mathcal{B}_{s,r}^{\NAV}(\tau)=
	\left\{
	(s,\NAV,k)\,\middle|\,
	k=1,\ldots,
	\min\{
	\widehat{R}_{s,r}^{\Tx}(\tau),
	\widehat{R}_{s,r}^{\Rx}(\tau)
	\}
	\right\}.
	\label{eq:nav_pair_copy_set}
\end{equation}

The matching graph is constructed according to the resource-conflict structure of each layer. For \(\ell_{\NAV}^{\mathrm{NS}}\) and \(\ell_{\NAV}^{\mathrm{NL}}\), a navigation-user node is connected to a paired satellite copy. The single user node allows at most one serving satellite to be selected for that user in one matching round, while different paired copies allow one multibeam satellite to serve several navigation users concurrently. For \(\ell_{\TM}^{\mathrm{U2S}}\), a TM-user node is connected to an Rx copy of a visible satellite, so each user selects at most one access satellite while a satellite may receive from several users through different Rx copies. For \(\ell_{\TC}^{\mathrm{D}}\), a satellite Tx copy is connected to the target TC-user node, allowing a satellite to deliver different events to multiple users through different Tx copies while preventing duplicate selection of the same target user in one round.

For the transfer layers
\(\ell_{\TM}^{\mathrm{N2A}}\),
\(\ell_{\TM}^{\mathrm{A2G}}\),
\(\ell_{\TC}^{\mathrm{G2A}}\), and
\(\ell_{\TC}^{\mathrm{A2N}}\),
each feasible source is represented by one source-state node and each destination by its available Rx copies. A candidate \(i\rightarrow j\) is represented by an edge between the source-state node of \(i\) and one Rx copy of \(j\). Hence, one source selects at most one destination in a single matching round, whereas one destination can receive from multiple sources through distinct Rx copies. The real Tx resource and phase of the source are still checked against the current ledger when matched edges are converted into directed actions and reserved.
Let
\[
G_{\ell_r,\tau}^{\mathrm{match}}
=
\left(
V_{\ell_r,\tau}^{\mathrm{match}},
E_{\ell_r,\tau}^{\mathrm{match}}
\right)
\]
be the beam-copy graph of layer \(\ell_r\). Each graph edge \(e_a\) represents one candidate action \(a\) and inherits \(w_a(\tau)\). The selected set is
\begin{equation}
	M_{\ell_r}(\tau)=
	\arg\max_{M\in\mathcal{M}(G_{\ell_r,\tau}^{\mathrm{match}})}
	\sum_{e_a\in M}w_a(\tau),
	\label{eq:beam_copy_matching_objective}
\end{equation}
where \(\mathcal{M}(G_{\ell_r,\tau}^{\mathrm{match}})\) is the feasible matching family. The copy structure prevents repeated use of the same beam resource while allowing one physical multibeam node to participate in multiple concurrent actions.

Two execution rules complete the Stage~2 procedure. First, TM is continuous, high-volume traffic, and its groundward relay can create substantial transfer pressure at non-anchor and anchor satellites. Under the source-state graph construction, one source can select only one destination in a single matching round, which may leave multiple Tx beams unused under heavy TM backlog. Therefore, the TM transfer layers
\(\ell_{\TM}^{\mathrm{N2A}}\) and
\(\ell_{\TM}^{\mathrm{A2G}}\)
may repeat candidate construction and matching for at most \(I_{\max}\) rounds. Across rounds, a layer-local working state removes TM items already reserved, allowing a high-pressure source to use additional Tx beams without repeatedly selecting the same TM batch. This working state is only for duplicate suppression and source-side transferable-amount updates; newly received TM still becomes available only after the half-slot boundary.

Second, navigation layers are executed only in the first half-slot of each slot. Once a navigation service is selected, its two opposite half-slot actions and corresponding satellite phases are reserved together; executing the navigation layers again in the second half-slot would duplicate the same slot-level ranging opportunity.

Algorithm~\ref{alg:dynamic_layer_scheduling} summarizes the complete Stage~2 execution in one half-slot. Layer-dependent urgency functions, exact marginal weights, filtering order, planned-item slicing, and post-matching verification are provided in the supplementary material.

\begin{algorithm}[t]
	\caption{Stage~2 Dynamic Layer Scheduling and Beam-Copy Matching}
	\label{alg:dynamic_layer_scheduling}
	\KwIn{Inputs inherited from Algorithm~\ref{alg:staged_half_slot_scheduler}; half-slot \((t,h)\); state snapshot \(\Omega_\tau\); layer order \(\mathcal{O}_\tau\); residual phase/resource ledger; \(\alpha_r,\beta_\ell\); TM matching limit \(I_{\max}\).}
	\KwOut{Selected Stage~2 actions in half-slot \(\tau\) and the updated resource ledger.}
	
	\If{\(h=1\)}{
		Remove \(\ell_{\NAV}^{\mathrm{NS}}\) and \(\ell_{\NAV}^{\mathrm{NL}}\) from \(\mathcal{O}_\tau\)\;
	}
	
	\For{\(r=1\) \KwTo \(|\mathcal{O}_\tau|\)}{
		Let \(\ell_r\) be the \(r\)-th layer in \(\mathcal{O}_\tau\)\;
		Compute \(\widehat{R}_{v,r}^{m}(\tau)\) for every node \(v\) and direction \(m\)\;
		Set \(N_r\leftarrow I_{\max}\) if
		\(\ell_r\in\{\ell_{\TM}^{\mathrm{N2A}},\ell_{\TM}^{\mathrm{A2G}}\}\);
		otherwise set \(N_r\leftarrow1\)\;
		
		\For{\(q=1\) \KwTo \(N_r\)}{
			Construct \(\mathcal{C}_{\ell_r}(\tau)\) from the current service state and layer semantics\;
			Filter candidates by visibility, phase/resource feasibility, layer caps, and positive marginal value\;
			\If{no candidate remains}{break\;}
			Build \(G_{\ell_r,\tau}^{\mathrm{match}}\) and solve the beam-copy maximum-weight matching\;
			\If{no action is selected}{break\;}
			Convert matched edges into directed actions, add them to \(\Pi\), update the real resource ledger, and reserve their planned service items\;
			\If{\(\ell_r\in\{\ell_{\TM}^{\mathrm{N2A}},\ell_{\TM}^{\mathrm{A2G}}\}\)}{
				Update the layer-local TM working state to suppress duplicate reservation\;
			}
		}
	}
	\Return selected actions and updated resource ledger\;
\end{algorithm}

\subsection{Half-Slot Reservation and State Commit}
\label{subsec:reservation_commit}

The scheduling procedure has a sequential algorithmic order but a parallel physical execution. Within one half-slot, Stage~1 actions reserve resources first, after which Stage~2 layers are processed according to the dynamic order. However, all selected actions in that half-slot are transmitted concurrently. The scheduler therefore separates \emph{reservation} from \emph{state commit}: reservation takes effect immediately for conflict prevention and duplicate suppression, whereas transmitted information becomes available only after the half-slot boundary.

When a directed action is selected, it is added to \(\Pi\), the corresponding endpoint phases are fixed, and one Tx beam at the transmitter and one Rx beam at the receiver are deducted from the resource ledger. If the action carries TM data or a TC event replica, its planned transmission items are also reserved in the current half-slot working state. Hence, later layers and later TM matching rounds observe the updated resource ledger and planned-item reservations, but still evaluate service feasibility from the state snapshot available at the beginning of the half-slot.

At the half-slot boundary, all planned service effects are committed simultaneously according to the transmission-completion model in Sec.~\ref{subsec:service_modeling}. TM arrivals become available at the receiving satellite or are recorded as delivered at a GS; TC copies update holder sets or complete events under the first-arrival rule; and navigation records are updated only after the required two-way ranging interaction is completed. Consequently, information transmitted by an earlier layer cannot be forwarded again by a later layer in the same half-slot, and a newly created TC replica cannot be further propagated until a subsequent half-slot. Exact planned-item, duplicate-suppression, boundary-update, and delay-recording rules are given in the supplementary material.
This reservation--commit separation prevents the algorithmic layer order from being interpreted as physical cascading within the same half-slot.

In summary, the proposed SDHSS framework trades the global joint optimality of \(P_0\) for an interpretable and scalable online scheduling structure. Stage~1 prevents GNSS baseline functions from being squeezed by large-scale service competition, while Stage~2 decomposes residual-resource allocation into state-driven local matching decisions. This structure allows the scheduler to operate with many users while preserving consistent service semantics, resource-conflict handling, and half-slot state evolution.

\section{Evaluation}
\label{sec:evaluation}

\subsection{Simulation Scenario and Parameter Settings}
\label{subsec:simulation_settings}

\paragraph*{a) Network and user scenario}
The simulation uses the BeiDou constellation, consisting of 24 MEO, 3 GEO, and 3 IGSO satellites~\cite{14}. Three GSs are located at Kashi, Weinan, and Sanya. The GS elevation mask is \(5^\circ\), while the PA-terminal scan ranges are \(60^\circ\) for MEO satellites and \(45^\circ\) for GEO/IGSO satellites; these constraints determine the FSA visibility indicator \(\chi_{ij}(f)\).

The external users comprise 50 short-range navigation users, 50 long-range navigation users, 100 TM users, and 50 TC users. Short-range navigation users are placed on GEO and IGSO orbits, whereas long-range users are placed near the Earth--Moon \(L_1\), \(L_3\), \(L_4\), and \(L_5\) regions and on lunar distant retrograde orbits~\cite{16}. The \(L_2\) region is excluded because of its poor visibility to GNSS satellites, and TM/TC users are placed on LEO orbits. These orbit assignments are used to generate representative visibility relationships rather than to restrict the scheduler to specific orbit classes. At the link-scheduling level, the scheduler observes only the FSA visibility graph. Once \(\chi_{ij}(f)\) is obtained, the continuous position of a user no longer enters the scheduling decision. Therefore, any user with valid visibility relationships to GNSS satellites can be served by the same scheduling framework regardless of its specific orbit.

\paragraph*{b) Time structure and Stage~1 settings}
The simulation spans one day, divided into 288 FSA states of 300~s each. Every FSA contains five 60-s superframes; one representative superframe is scheduled, and its schedule is reused by the remaining four superframes because the visibility graph is fixed within the FSA. Each superframe contains 20 slots of 3~s, and each slot is divided into two 1.5-s half-slots. The superframe is also the evaluation window for GNSS ranging diversity and navigation-user ranging satisfaction. Stage~1 uses \(L_{\min}=10\) distinct ranging counterparts per satellite and a sliding-window length \(T_m=3\) slots for baseline TM/TC access protection.

\paragraph*{c) Beam resources and link capacities}
The satellite beam numbers \(X_s\) and \(Y_s\) are experimental variables used to evaluate the transition from conventional single-beam operation to multibeam operation. Unless otherwise stated, each GS has 4 Tx beams and 4 Rx beams. Directed-link capacities, expressed in the same data unit as TM batches and virtual TC events, are set to 20 for user--satellite links, 60 for inter-satellite links, 200 for satellite-to-GS links, and 60 for GS-to-satellite links.

\paragraph*{d) Traffic generation and navigation demand}
TM is modeled as real periodic traffic. Each GNSS satellite generates one TM batch in every slot, and each TM user generates one TM batch every two slots. The amount of each batch is 10. This setting creates a high-frequency TM workload that exposes the throughput pressure and backlog evolution of groundward TM return.

TC is modeled through virtual event probes. Since a real TC may be issued at an arbitrary time, both satellite-targeted and user-targeted TC probes are generated in every slot. The completion delay of these probes measures the low-delay reachability that the schedule can provide to real TC events released at different times. Each TC probe has amount 0.1, much smaller than the TM batch size, because the probes represent small, highly addressed command events rather than sustained bulk traffic. Their purpose is therefore to evaluate reachability and delivery delay rather than throughput.

For navigation users, following prior studies~\cite{3,22}, each user is assumed to require four distinct valid ranging measurements per superframe.

\paragraph*{e) Stage~2 preference parameters}
The numerical coefficients for dynamic layer ordering, layer resource sharing, iterative TM transfer matching, service priorities, aging, and marginal link weighting are treated as preference-tuning parameters. Their exact values and tuning effects are reported in the supplementary material.

\subsection{Baseline Method}
\label{subsec:baseline_method}

We use Fair Contact Plan (FCP)~\cite{12} as the comparison baseline. Existing GNSS link-scheduling methods mainly address intra-constellation contact planning and do not directly specify large-scale user service allocation. FCP is suitable for comparison because its fairness mechanism can operate on a unified graph containing satellite--satellite, satellite--GS, and satellite--user contacts. It nevertheless retains the conventional slot-level one-\(\Tx\) one-\(\Rx\) bidirectional contact model, in which each satellite or user communicates with at most one counterpart per slot.
For FSA state \(f\), FCP builds an undirected candidate graph \(\mathcal{G}^{\mathrm{FCP}}_f\) from all visible satellite--satellite, satellite--GS, and satellite--user contacts. To represent GS multibeam capability without changing the one-contact-per-satellite assumption, each GS \(g\) is expanded into
\[
B_g^{\mathrm{pair}}=\min\{X_g,Y_g\}
\]
virtual copies, whereas satellites and users remain unsplit. A GS may therefore support multiple simultaneous contacts, while each satellite and user is incident to at most one selected contact per slot.
Let \(q_e(t)\) be the fairness weight of edge \(e\) in slot \(t\), initialized by \(q_e(0)=1\). FCP selects
\begin{equation}
	M^{\mathrm{FCP}}(t)=
	\arg\max_{M\in\mathcal{M}(\mathcal{G}^{\mathrm{FCP}}_f)}
	\sum_{e\in M}q_e(t),
	\label{eq:fcp_matching}
\end{equation}
and updates
\begin{equation}
	q_e(t+1)=
	\begin{cases}
		1, & e\in M^{\mathrm{FCP}}(t),\\
		q_e(t)+\Delta_{\kappa(e)}, & e\notin M^{\mathrm{FCP}}(t),
	\end{cases}
	\label{eq:fcp_weight_update}
\end{equation}
where \(\mathcal{M}(\mathcal{G}^{\mathrm{FCP}}_f)\) is the feasible matching family and \(\kappa(e)\) denotes the edge type. We set
\[
\Delta_{\mathrm{SS}}=4,\qquad
\Delta_{\mathrm{SG}}=2,\qquad
\Delta_{\mathrm{SU}}=1.5,
\]
so that the much larger number of user-related candidates does not overwhelm GNSS-internal and ground-access contacts.

For consistent metric evaluation, each selected FCP contact is expanded into two opposite directed half-slot actions within the same slot. The actions follow the same ranging, TM, TC, and half-slot completion semantics used for SDHSS and contribute only when their directions and endpoints produce valid service transitions under the current state. However, both half-slot directions remain tied to the same counterpart. FCP therefore retains slot-level bidirectional contact planning, whereas SDHSS can schedule service-driven directed actions independently at half-slot granularity.

\subsection{Evaluation Results}
\label{subsec:evaluation_results}

We evaluate six configurations. SDHSS is tested with 1\(\Tx\)/1\(\Rx\) PA antenna both without and with users, and with 2\(\Tx\)/2\(\Rx\) and 3\(\Tx\)/3\(\Rx\) in the user scenario. FCP is evaluated with 1\(\Tx\)/1\(\Rx\), both without and with users. The no-user cases examine compatibility with constellation-internal operation, whereas the user cases evaluate open-GNSS service under limited and expanded beam resources.

For TM and TC, end-to-end delay is measured in half-slot units from source generation to destination arrival. Under the half-slot completion model, the minimum delay is one half-slot for an anchor satellite (involving a single hop interaction with the GS) and two for a non-anchor satellite (requiring two hops via the anchor satellite and the GS); for a user served through an anchor or non-anchor satellite, the corresponding minima are two and three half-slots, respectively. 

The delivery completion ratio is the successfully delivered original data amount divided by the generated original amount. TM is counted as delivered only after reaching a GS. For TC, the first replica reaching the target completes the event, and the original event amount is counted once regardless of how many replicas were created.

\subsubsection{Runtime and overall feasibility}
\label{subsubsec:runtime_feasibility}

\begin{figure}[t]
	\centering
	\includegraphics[width=0.93\linewidth]{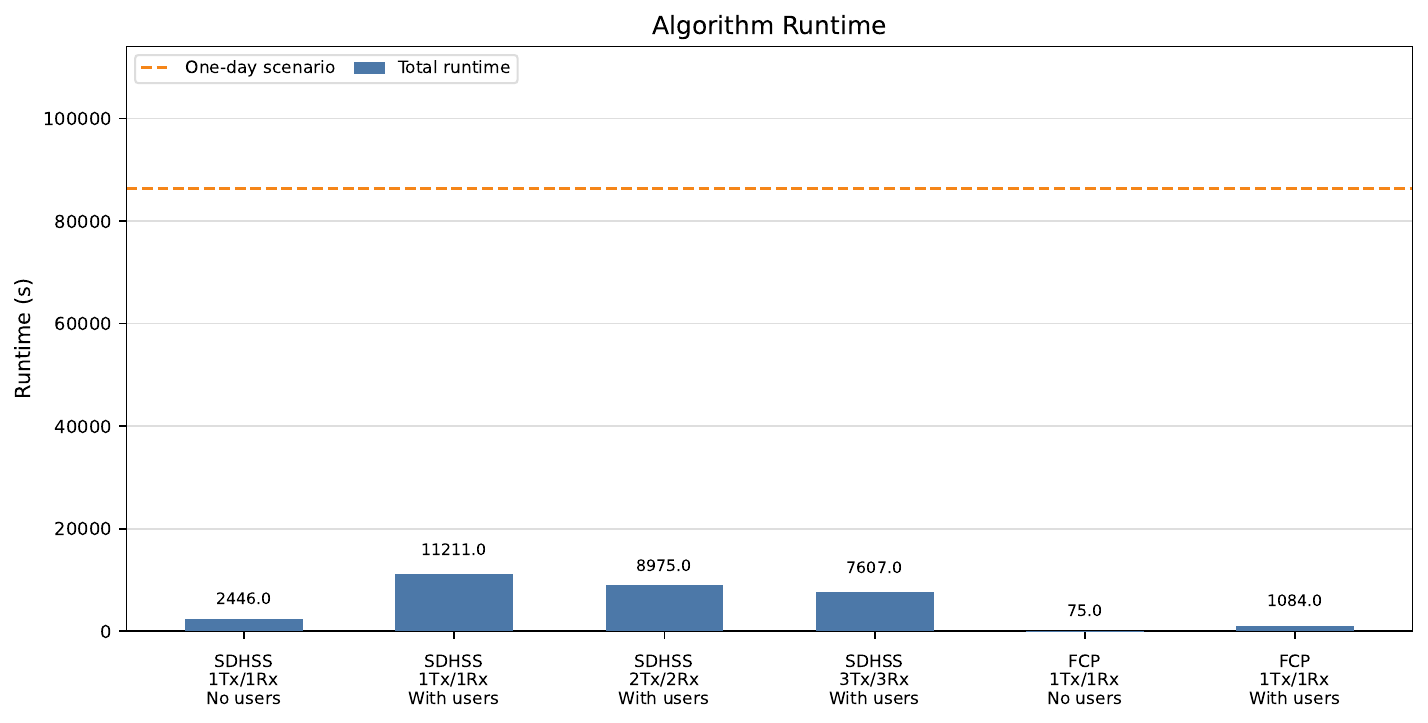}
	\caption{Runtime comparison under different scheduling configurations.}
	\label{fig:algorithm_runtime}
\end{figure}

Fig.~\ref{fig:algorithm_runtime} reports the runtime of the one-day simulation on an AMD Ryzen 7 7700X 8-Core Processor with 32~GB RAM. SDHSS requires more computation than FCP because it schedules half-slot actions, maintains service states, and repeatedly performs layer-wise matching. Nevertheless, every SDHSS configuration completes well within the 86400-s simulated horizon; the longest runtime is about \(1.12\times10^4\)~s for the 1\(\Tx\)/1\(\Rx\) user case.

The runtime is not monotonic in the beam number. Under scarce 1\(\Tx\)/1\(\Rx\) resources, TM backlogs and unfinished TC events persist longer, enlarging the state ledger and repeated candidate checks. Additional beams enlarge the instantaneous action space but also drain TM and complete TC events faster, reducing later scheduling-state complexity. 
Overall, the results therefore demonstrate computational feasibility for the evaluated open-GNSS scale.

\subsubsection{GNSS baseline operation}
\label{subsubsec:gnss_baseline_operation}

\begin{figure}[t]
	\centering
	\includegraphics[width=0.98\linewidth]{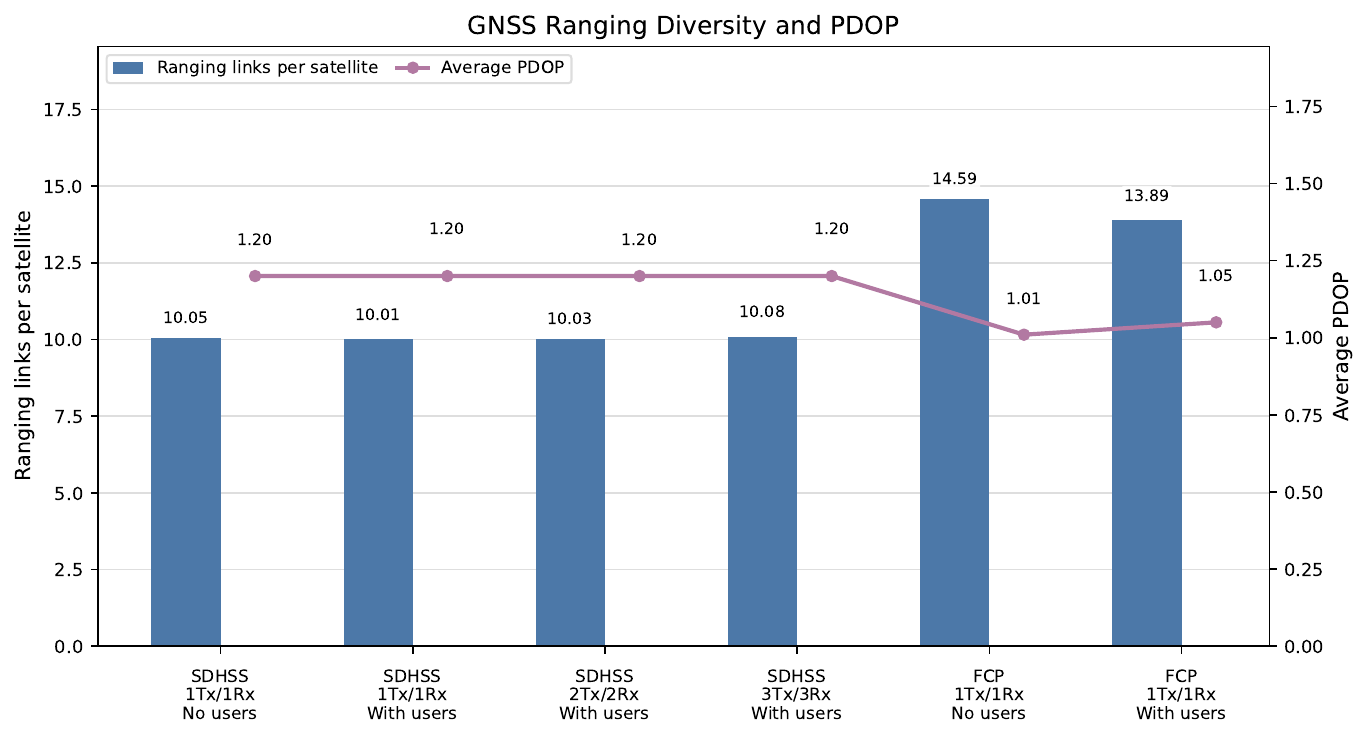}
	\caption{GNSS ranging diversity and average PDOP.}
	\label{fig:ranging_diversity_pdop}
\end{figure}

Fig.~\ref{fig:ranging_diversity_pdop} shows the average number of ranging links per GNSS satellite and the corresponding average PDOP. The proposed scheduler keeps the average ranging diversity slightly above ten in all tested configurations, and the average PDOP stays around 1.20. 
Hence, adding large-scale users or additional service traffic does not degrade the Stage~1-protected ranging baseline. FCP obtains more ranging contacts and lower PDOP in the 1\(\Tx\)/1\(\Rx\) cases because it continues to spend slot-level bidirectional contacts on satellite--satellite and satellite--GS edges. SDHSS instead preserves the required ranging baseline and leaves additional opportunities for communication and user services.

\begin{figure}[t]
	\centering
	\includegraphics[width=0.95\linewidth]{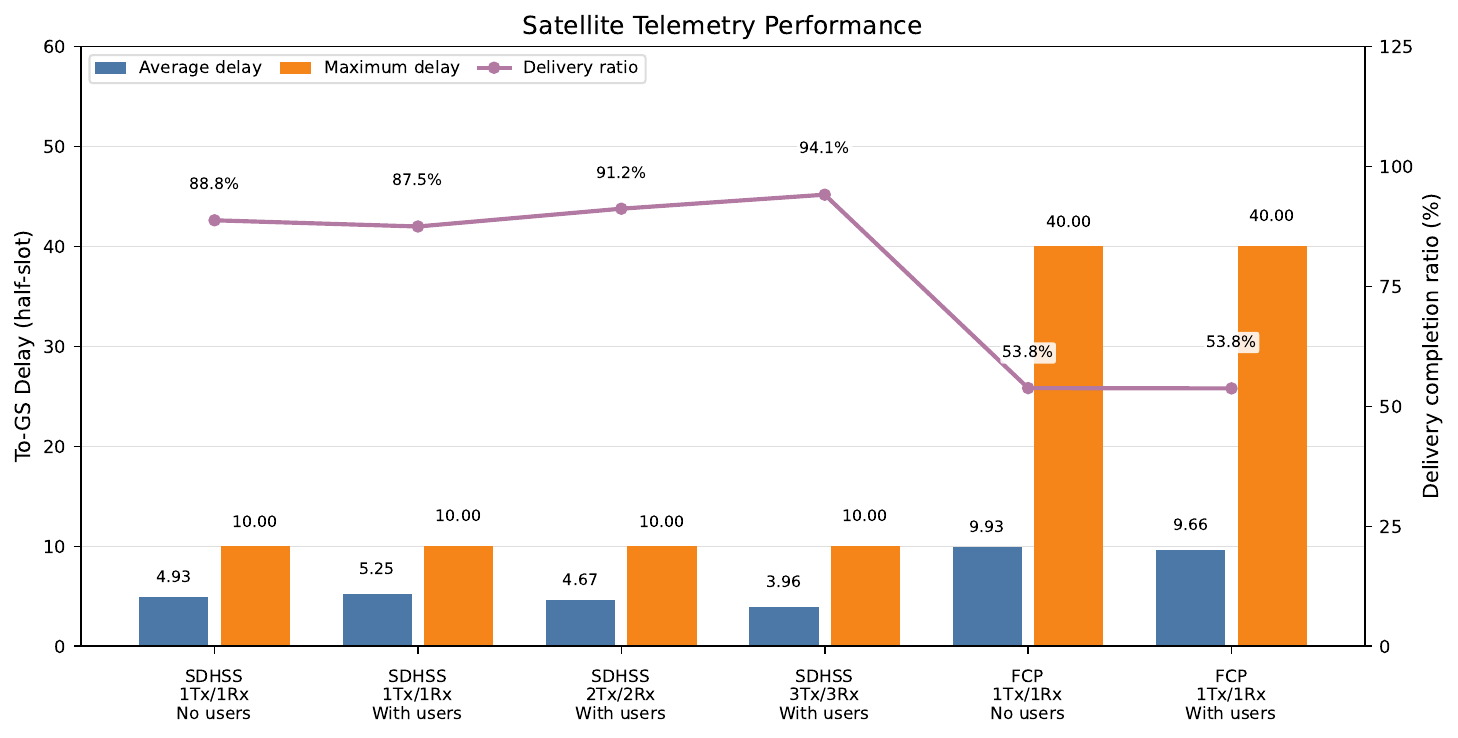}
	\caption{GNSS-satellite telemetry delivery performance.}
	\label{fig:satellite_telemetry_performance}
\end{figure}

\begin{figure}[t]
	\centering
	\includegraphics[width=0.95\linewidth]{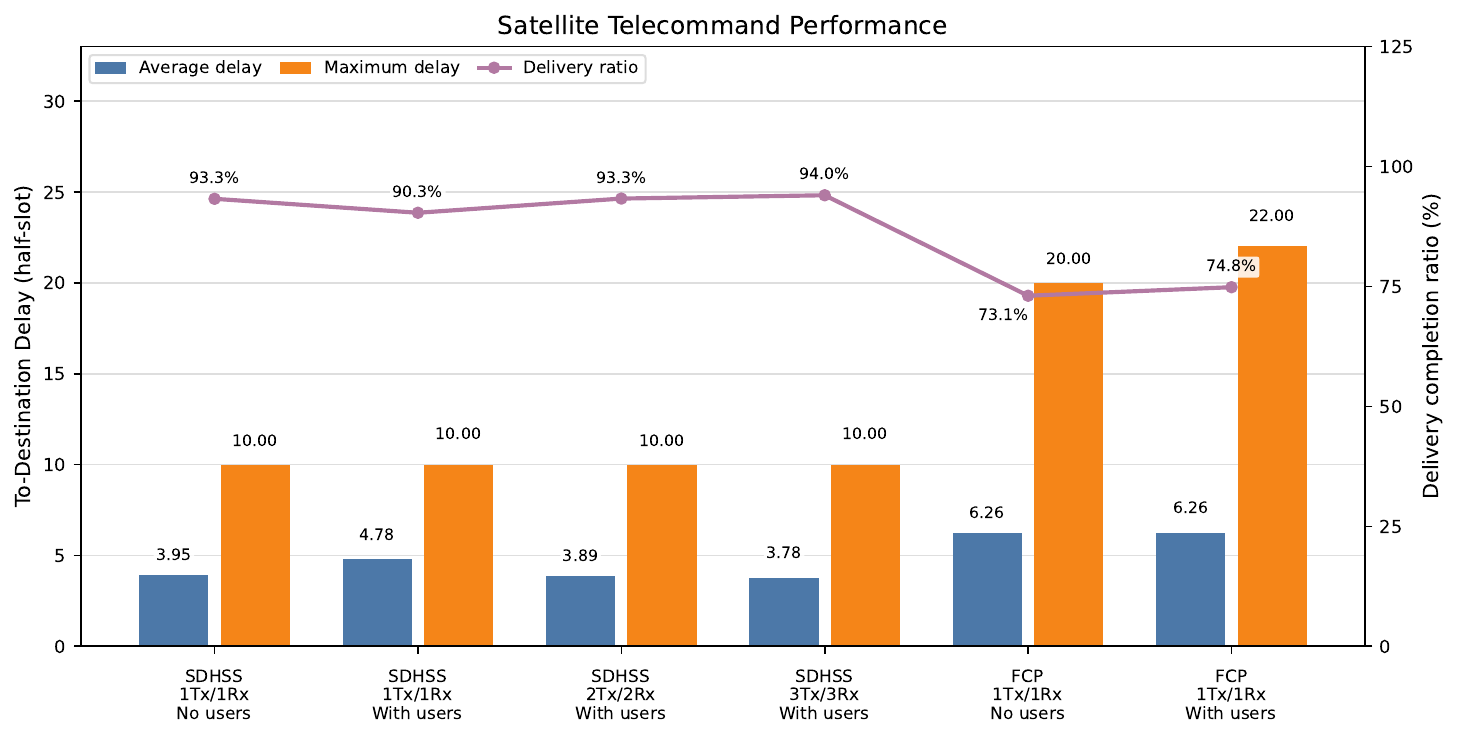}
	\caption{GNSS-satellite telecommand delivery performance.}
	\label{fig:satellite_telecommand_performance}
\end{figure}

Figs.~\ref{fig:satellite_telemetry_performance} and~\ref{fig:satellite_telecommand_performance} show that baseline protection also extends to GNSS-satellite communication. For satellite TM, SDHSS keeps the average delay within about five half-slots and the maximum delay at ten half-slots in all tested configurations; the delivery ratio reaches 94.1\% with 3\(\Tx\)/3\(\Rx\). FCP delivers only 53.8\% of satellite TM in both 1\(\Tx\)/1\(\Rx\) cases, with a maximum delay of 40 half-slots. The gap shows that fair contact occurrence alone does not ensure causal groundward progress of queued TM.

For satellite-targeted TC, SDHSS achieves completion ratios of 90.3--94.0\%, average delays of 3.78--4.78 half-slots, and a maximum delay of ten half-slots. FCP reaches only 73.1--74.8\% completion, with average delay 6.26 half-slots and maximum delays of 20--22 half-slots. The difference reflects the event-state-aware TC mechanism: SDHSS values links that inject, expand, or complete an unfinished event, whereas FCP schedules contacts according to fairness weights without considering the current replica-holder state.

Completion ratios below 100\% are also affected by the finite 20-slot evaluation window. TM batches and TC probes generated near the superframe boundary may require several half-slots to traverse a source-to-destination path and therefore may not be completed before the superframe boundary. This is a within-superframe measurement effect rather than a loss of the data or event in continuous operation.

\subsubsection{User service performance}
\label{subsubsec:external_user_service_results}

\begin{figure}[t]
	\centering
	\includegraphics[width=0.93\linewidth]{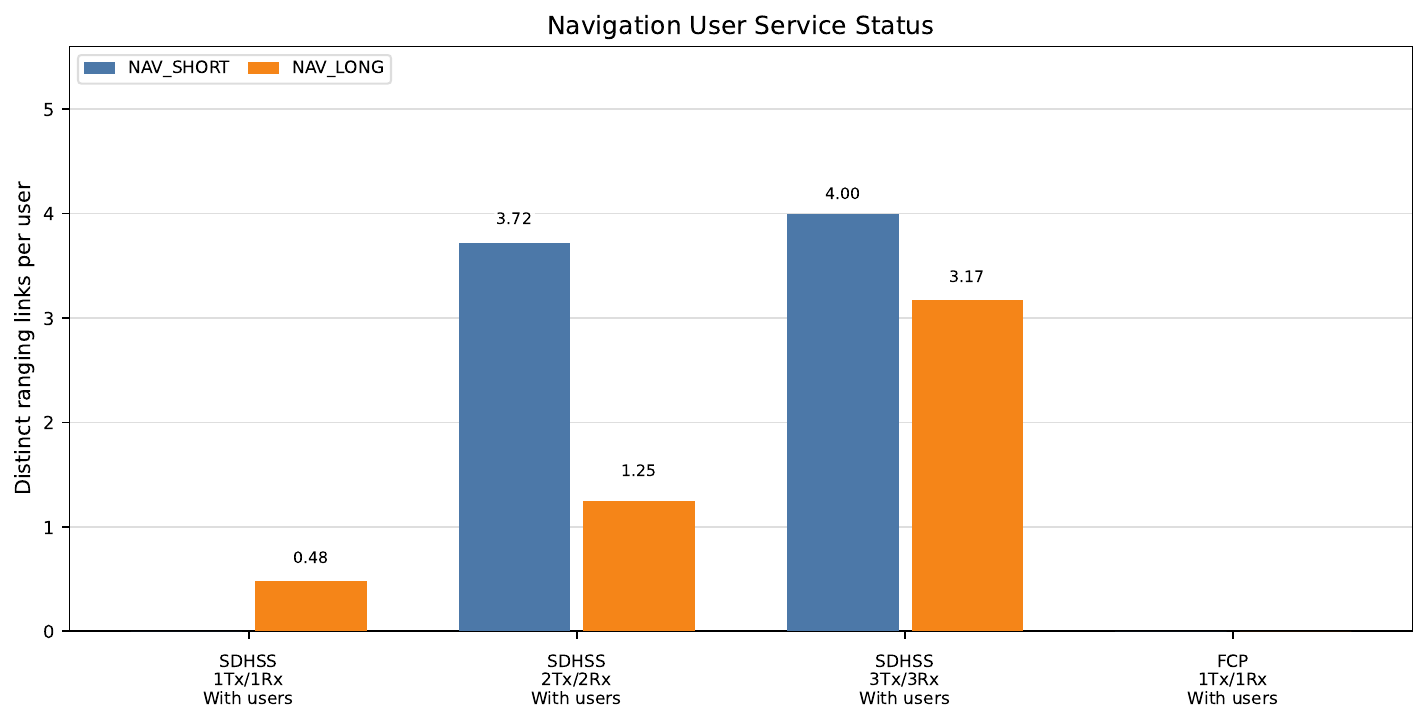}
	\caption{Distinct ranging links obtained by navigation users.}
	\label{fig:navigation_user_service}
\end{figure}

Fig.~\ref{fig:navigation_user_service} shows that the current single-beam payload leaves little capacity for navigation users under multi-service competition. With 1\(\Tx\)/1\(\Rx\), short-range users obtain almost no service, while long-range users obtain only 0.48 distinct serving satellites on average. Additional beams rapidly release navigation capacity: short-range users obtain 3.72 distinct satellites with 2\(\Tx\)/2\(\Rx\) and reach the target value of four with 3\(\Tx\)/3\(\Rx\); long-range users reach 3.17 with 3\(\Tx\)/3\(\Rx\). Their lower performance is caused by the stricter \(\RT\) phase requirement, confirming that long-range navigation depends on phase-compatible opportunities rather than visibility alone.

\begin{figure}[t]
	\centering
	\includegraphics[width=0.95\linewidth]{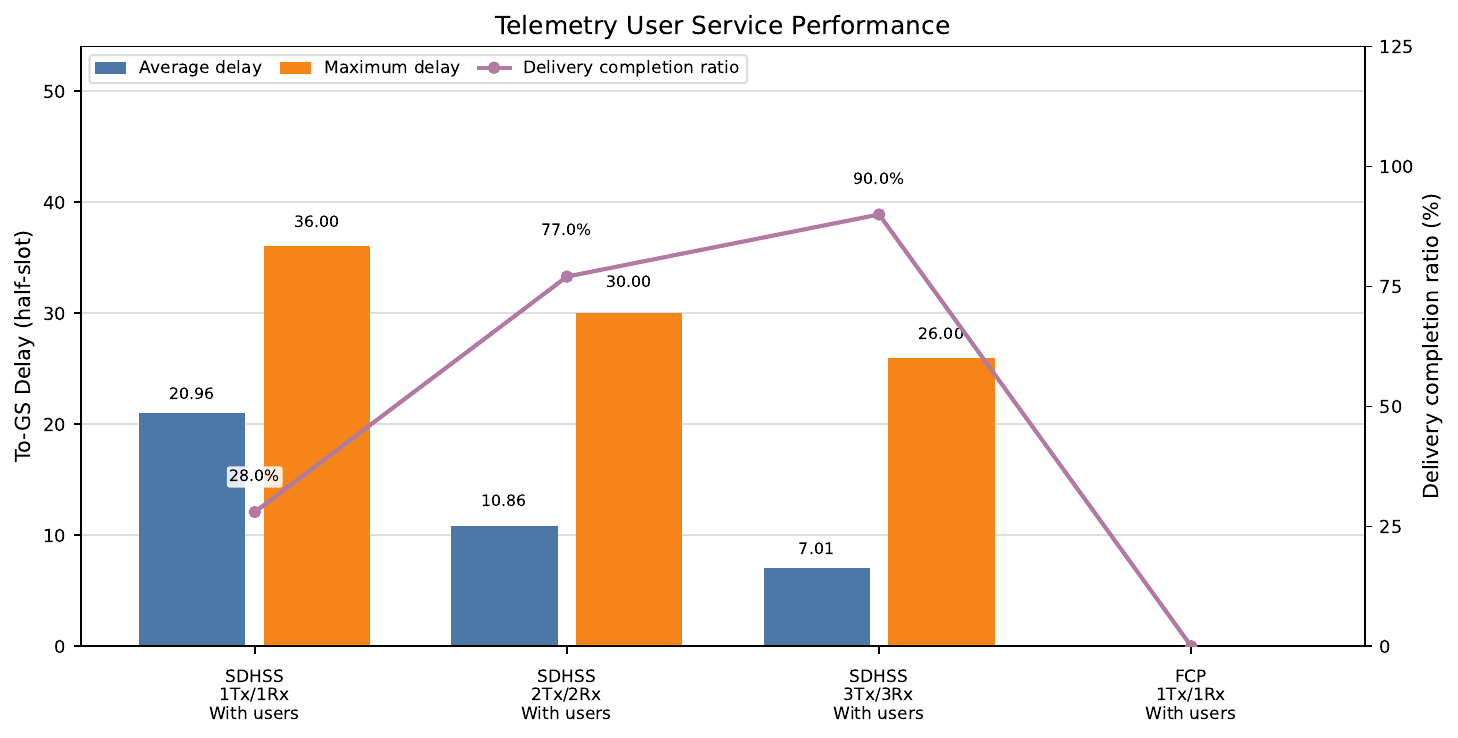}
	\caption{Telemetry user delivery performance.}
	\label{fig:telemetry_user_service}
\end{figure}

\begin{figure}[t]
	\centering
	\includegraphics[width=0.95\linewidth]{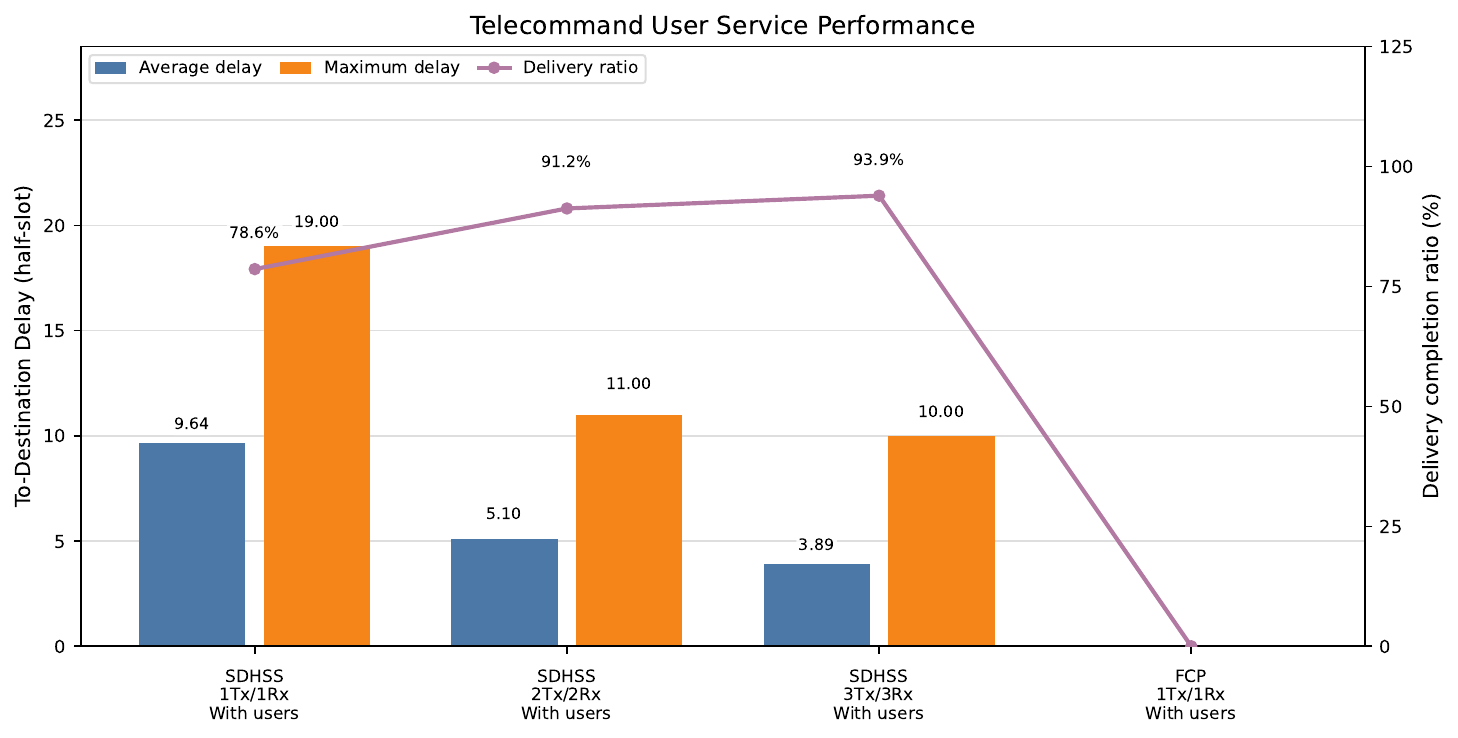}
	\caption{Telecommand user delivery performance.}
	\label{fig:telecommand_user_service}
\end{figure}

Figs.~\ref{fig:telemetry_user_service} and~\ref{fig:telecommand_user_service} show the same beam-resource effect for user communication. For TM users, increasing the satellite configuration from 1\(\Tx\)/1\(\Rx\) to 3\(\Tx\)/3\(\Rx\) raises the delivery ratio from 28.0\% to 90.0\% and reduces the average delay from 20.96 to 7.01 half-slots. Additional Rx beams admit more user TM, while additional Tx beams increase non-anchor-to-anchor and anchor-to-GS transfer opportunities.

For TC users, the completion ratio rises from 78.6\% with 1\(\Tx\)/1\(\Rx\) to 93.9\% with 3\(\Tx\)/3\(\Rx\), where the average delay is 3.89 half-slots. TC remains faster than TM because the probes are small and completion-oriented, whereas periodic TM creates sustained throughput pressure. User performance is generally lower than GNSS-satellite performance, which is consistent with the design objective. GNSS must first maintain its own ranging, TM return, and TC reachability. User data are relayed by the GNSS constellation and have lower service priority in normal conditions. However, the age-dependent urgency and marginal weights prevent user data from being permanently starved: old TM batches and delayed TC events become increasingly competitive and can obtain resources when they have waited for a long time.

\subsubsection{Link composition and service-driven asymmetric topology}
\label{subsubsec:link_composition_asymmetry}

\begin{figure}[t]
	\centering
	\includegraphics[width=0.48\textwidth]{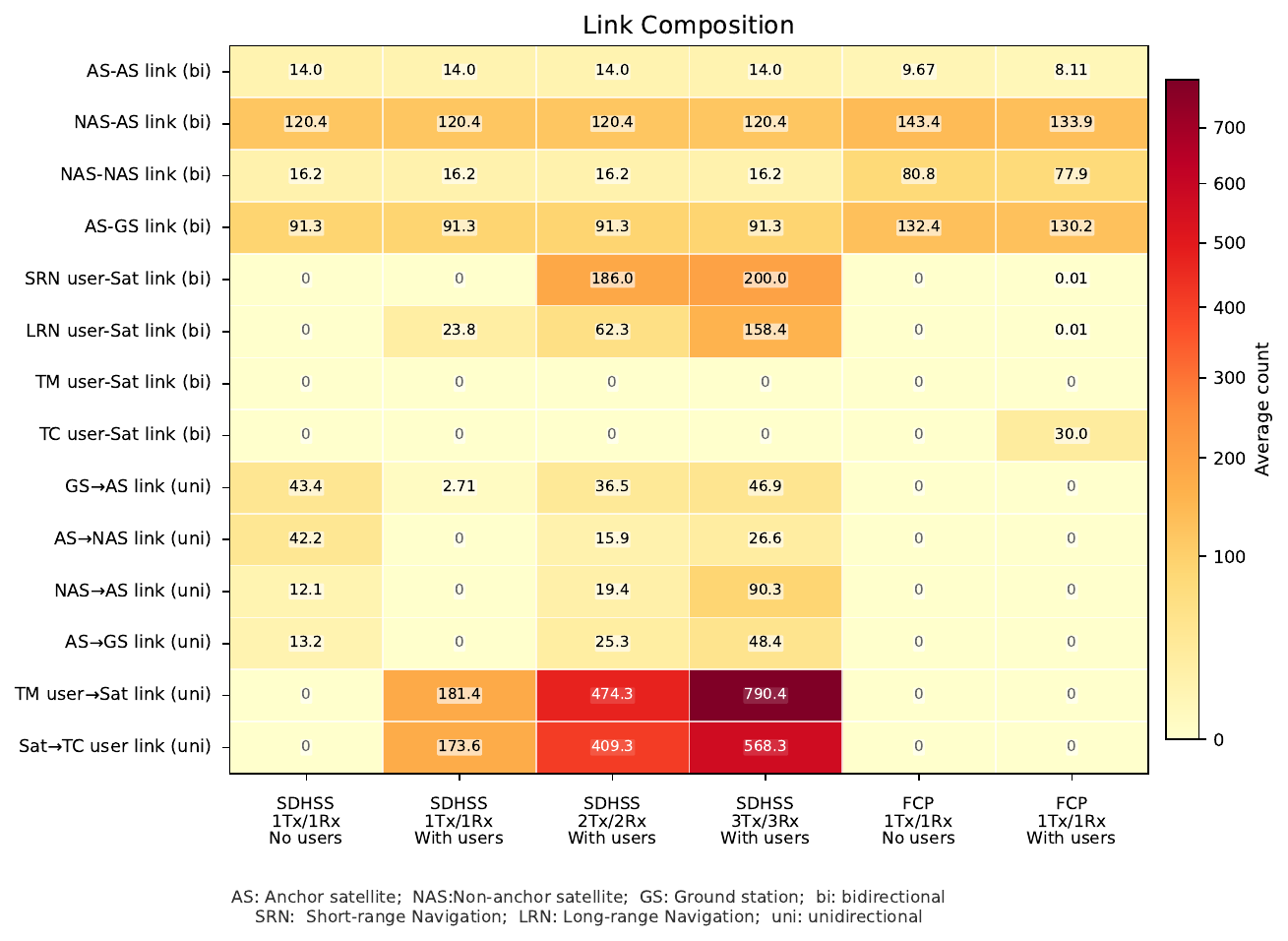}
	\caption{Average link composition under different scheduling configurations.}
	\label{fig:link_composition_heatmap}
\end{figure}
Fig.~\ref{fig:link_composition_heatmap} explains the preceding performance trends through the scheduled link composition. Stage~1 bidirectional satellite--satellite and satellite--GS contacts remain relatively stable across SDHSS beam configurations because they protect the GNSS baseline. The main changes occur in Stage~2 directed transfer and user-service links, whose composition adapts to the service state and available beams.

With 1\(\Tx\)/1\(\Rx\), scarce resources make delay-sensitive TC injection and dissemination highly competitive, so upward TC-related actions are comparable to or more frequent than downward TM-transfer actions. As the beam number increases, the sustained TM throughput demand can be served more fully: non-anchor-to-anchor and anchor-to-GS actions increase substantially, and the downlink direction becomes dominant with 3\(\Tx\)/3\(\Rx\). The topology therefore becomes asymmetric according to service demand, delay sensitivity, and available resources rather than following a fixed uplink/downlink symmetrical balance.

FCP cannot produce this adaptation because every selected contact binds both half-slots to the same counterpart. More fundamentally, fairness over visible contacts does not guarantee useful service transitions: TM links must advance stored data toward a GS, while TC propagation requires the transmitter to hold the corresponding unfinished event and the receiver to provide useful propagation or completion value. SDHSS evaluates these states before link selection, so additional beams are converted into concurrent service-effective transitions rather than merely more visible contacts. This causal link between beam resources, directed link composition, and service performance explains the gains observed above.

\section{Conclusion}
\label{sec:conclusion}

This paper addressed half-slot link scheduling for an open GNSS equipped with \(x\)-Tx \(y\)-Rx multibeam phased-array antennas. The problem is to preserve GNSS ranging and communication functions while using rapidly reconfigurable beam resources to support large-scale navigation, TM, and TC users. Compared with conventional slot-level bidirectional contact planning, this setting requires directed half-slot scheduling, explicit Tx/Rx phase control, multibeam concurrency management, and service-aware link selection.

To address these requirements, we developed SDHSS, which combines differentiated NAV/TM/TC service models with staged resource allocation. Stage~1 reserves a sparse baseline contact template for GNSS ranging diversity and recurring contact opportunities for satellite TM return and TC reachability, while Stage~2 dynamically allocates the remaining half-slot resources according to navigation deficits, TM backlogs, and TC event states. Dynamic service-layer ordering and beam-copy maximum-weight matching jointly adapt resource competition to current demand while preserving multibeam concurrency and resource feasibility.

BeiDou-based simulations show that SDHSS preserves the GNSS baseline while providing substantially stronger communication and external-user service performance than FCP. As satellite beam resources increase, additional Tx/Rx beams are effectively converted into greater navigation ranging support, improved TM delivery, and faster TC completion. The resulting Stage~2 topology is service-driven and directionally asymmetric, reflecting the different throughput and delay requirements of TM and TC rather than forcing all contacts into slot-level bidirectional pairs. These results show that half-slot, state-aware multibeam scheduling provides a suitable scheduling basis for extending GNSS from constellation-internal operation toward open space infrastructure.

Future work will investigate coordination with laser backbone links~\cite{18} and more detailed mission-specific traffic models for heterogeneous spacecraft users.

\bibliographystyle{IEEEtran}
\bibliography{references}


	\section{Supplementary Material: Operational Details for Staged Half-Slot Scheduling}
	\label{sec:supp_operational_details}
	
	The main paper defines the scheduling architecture and the design logic of SDHSS. This section specifies the runtime state, exact urgency and marginal-value calculations, item reservation rules, matching execution, and half-slot boundary updates needed for an implementation-equivalent realization. The notation follows the main paper: \(\tau=2t+h\) is the ordered half-slot index, \(\chi_{ij}(f)\) is the FSA visibility indicator, \((\Aset_f,\Nset_f)\) is the anchor/non-anchor partition in FSA state \(f\), \(R_v^{\Tx}(\tau)\) and \(R_v^{\Rx}(\tau)\) are the current residual beam resources, and \(\Omega_\tau\) is the physical service-state snapshot available at the beginning of half-slot \(\tau\). Table~\ref{tab:supp_correspondence} maps the main-paper mechanisms to the corresponding runtime details in this supplement.
	
	\begin{table*}[t]
		\centering
		\caption{Correspondence between main-paper mechanisms and supplementary runtime details.}
		\label{tab:supp_correspondence}
		\footnotesize
		\begin{tabular}{p{0.28\textwidth}p{0.42\textwidth}p{0.20\textwidth}}
			\toprule
			Main-paper mechanism & Runtime detail specified in this supplement & Location \\
			\midrule
			Runtime state and resource control &
			physical service snapshot, working reservations, phase state, and residual-resource ledger &
			Sec.~\ref{subsec:supp_runtime_state} \\
			Stage~1 oriented actions &
			carried-service item selection and direction scoring for same-role satellite contacts &
			Sec.~\ref{subsec:supp_stage1_actions} \\
			Stage~2 dynamic ordering &
			exact urgency functions for navigation, TM, and TC layers &
			Sec.~\ref{subsec:supp_urgency} \\
			Candidate construction and weighting &
			filtering order, item queues, marginal weights, and capacity-sliced planned items &
			Sec.~\ref{subsec:supp_candidates} \\
			Beam-copy matching &
			copy-level graph expansion, post-matching verification, navigation reservation, and iterative TM execution &
			Sec.~\ref{subsec:supp_matching} \\
			Reservation and state commit &
			duplicate suppression, exact boundary updates, first-arrival semantics, and delay recording &
			Sec.~\ref{subsec:supp_commit} \\
			Stage~2 preference parameters &
			numerical settings and tuning rules for ordering, resource sharing, aging, and TC completion/frontier gains &
			Sec.~\ref{sec:supp_preference_parameters} \\
			\bottomrule
		\end{tabular}
	\end{table*}

	\subsection{Runtime State and Resource Ledger}
	\label{subsec:supp_runtime_state}
	
	At the beginning of half-slot \(\tau\), the scheduler observes the physical service-state snapshot \(\Omega_\tau\). This snapshot contains only information that is available before transmissions in half-slot \(\tau\) start. For navigation user \(u\), \(Q_u(\tau)\) is the set of GNSS satellites whose required two-way ranging interactions have already been completed and counted in the current superframe. The TM state consists of the user-side backlog \(B_u^{\TM}(\tau)\) for each \(u\in\Uset^{\mathrm{M}}\) and the satellite-side TM batch set \(\mathcal{B}_s^{\TM}(\tau)\) for each \(s\in\Sset\). The TC state consists of the unfinished-event set \(\Eset_{\mathrm{open}}^{\TC}(\tau)\). Each event \(e\in\Eset_{\mathrm{open}}^{\TC}(\tau)\) has generation slot \(g_e\), amount \(A_e\), destination \(d(e)\), and holder set \(H_e(\tau)\).
	
	At the beginning of each independently evaluated representative superframe, the service state is freshly initialized:
	\[
	Q_u(0)=\varnothing,\qquad
	B_u^{\TM}(0)=0,\qquad
	\mathcal{B}_s^{\TM}(0)=\varnothing,
	\]
	and no unfinished TC event is carried over from another representative superframe. TM batches and TC events are then inserted according to the slot-level generation rules used in the evaluation. A TM batch \(b\) generated in slot \(g_b\) becomes source-available at boundary \(2g_b\). Likewise, a TC event \(e\) generated in slot \(g_e\) becomes available at boundary \(2g_e\) and is initially held by all GSs:
	\begin{equation}
		H_e(2g_e)=\Gset.
		\label{eq:supp_tc_initial_holders}
	\end{equation}
	
	The scheduler also maintains a resource ledger and a runtime phase state. At the beginning of a half-slot, each node starts from its physical beam capacities minus any action already reserved for that half-slot, including Stage~1 actions and the second-half action of a navigation service selected in the first half-slot. The resulting counters are \(R_v^{\Tx}(\tau)\) and \(R_v^{\Rx}(\tau)\).
	
	To distinguish the runtime phase ledger from the binary phase variable \(\theta_s(t,h)\) used in the main paper, let
	\[
	\varphi_s(\tau)\in\{\Tx,\Rx,\UNK\}
	\]
	denote the current runtime phase state of satellite \(s\), where \(\UNK\) means that no accepted or pre-reserved action has yet fixed the mode of that satellite in half-slot \(\tau\). The two half-slots of one slot must remain complementary. Therefore, fixing
	\(\varphi_s(2t)=\Tx\) also fixes \(\varphi_s(2t+1)=\Rx\), and fixing
	\(\varphi_s(2t)=\Rx\) fixes \(\varphi_s(2t+1)=\Tx\); the same rule applies in reverse if the second-half mode has already been pre-reserved. Any candidate inconsistent with an already fixed complementary phase is rejected.
	
	A directed action \(a:i\rightarrow j\) in half-slot \(\tau\) is physically feasible only if
	\begin{align}
		\chi_{ij}(f)&=1,
		\label{eq:supp_visibility}\\
		R_i^{\Tx}(\tau)&>0,\qquad
		R_j^{\Rx}(\tau)>0,
		\label{eq:supp_resource}\\
		i\in\Sset&\Rightarrow
		\varphi_i(\tau)\in\{\Tx,\UNK\},
		\nonumber\\
		j\in\Sset&\Rightarrow
		\varphi_j(\tau)\in\{\Rx,\UNK\}.
		\label{eq:supp_phase}
	\end{align}
	User endpoints additionally obey the service-layer direction and single-counterpart constraint defined in the main paper. GS Tx and Rx resources are independent; a GS may therefore transmit on some actions and receive on others in the same half-slot as long as the respective residual counters remain positive.
	
	Once an action is accepted, one Tx beam is deducted from its transmitter and one Rx beam from its receiver. If a satellite endpoint was previously \(\UNK\), its mode in the current half-slot and the complementary mode in the other half-slot of the same slot are fixed. These resource and phase updates are immediate and are visible to later Stage~2 layers. In contrast, TM arrivals, TC holder-set changes, event completions, and navigation completion records are not written into \(\Omega_\tau\) during the half-slot. Candidate construction therefore uses the fixed physical snapshot \(\Omega_\tau\) together with two runtime overlays: the updated resource/phase ledger and the set of service items already reserved earlier in the same half-slot.
	
	For the layer \(\ell_r\) executed at rank \(r\), the implementation uses the per-layer beam cap defined in the main paper. Equivalently, define the nominal cap
	\begin{equation}
		\bar{R}_{v,r}^{m}
		=
		\max\left\{
		1,
		\left\lfloor
		\eta_{\ell_r,r}C_v^m
		\right\rfloor
		\right\},
		\qquad
		m\in\{\Tx,\Rx\},
		\label{eq:supp_nominal_layer_cap}
	\end{equation}
	where \(C_v^{\Tx}=X_v\), \(C_v^{\Rx}=Y_v\), and
	\(\eta_{\ell_r,r}=\min\{\alpha_r,\beta_{\ell_r}\}\).
	The number of resource copies exposed to the current layer is
	\begin{equation}
		\widehat{R}_{v,r}^{m}(\tau)=
		\begin{cases}
			\min\{R_v^m(\tau),\bar{R}_{v,r}^{m}\},
			&
			R_v^m(\tau)>0,\ \eta_{\ell_r,r}>0,\\
			0, & \text{otherwise}.
		\end{cases}
		\label{eq:supp_layer_resource_cap}
	\end{equation}
	The cap applies to the entire execution of the current layer, and all matching rounds of an iterative TM transfer layer share this cumulative cap. The physical residual ledger \(R_v^m(\tau)\) is reduced only after matched actions pass the final feasibility check and are reserved. Numerical values of \(\alpha_r\), \(\beta_\ell\), and \(I_{\max}\) are given in Sec.~\ref{subsec:supp_resource_params}.

	\subsection{Stage~1 Oriented Actions and Carried-Service Effects}
	\label{subsec:supp_stage1_actions}
	
	Stage~1 selects slot-level bidirectional baseline contacts, while the main paper specifies how each selected contact is oriented into two opposite half-slot actions. Once oriented, a Stage~1 action enters the same physical resource ledger as Stage~2 actions. For an action executed in half-slot \(\tau\), the carried-service effect is evaluated from \(\Omega_\tau\), excluding items already reserved earlier in the same half-slot, and the selected items are committed only at boundary \(\tau+1\).
	
	The service that can be carried by an oriented Stage~1 action follows the same routing semantics as Stage~2. In particular,
	\[
	g\rightarrow a
	\ \text{and}\
	a\rightarrow n
	\]
	can carry useful TC event replicas, whereas
	\[
	n\rightarrow a
	\ \text{and}\
	a\rightarrow g
	\]
	can carry TM items. Stage~1 anchor--anchor and non-anchor--non-anchor actions do not carry TM because same-role TM forwarding is excluded by the TM model; they may, however, propagate useful TC replicas. Thus, the direction and endpoint roles determine the eligible service queue, avoiding an artificial division of traffic by scheduling stage.
	
	For a TM-carrying Stage~1 action \(a:i\rightarrow j\) with capacity \(c_a\), the scheduler forms the eligible TM queue using the same batch-value ordering and capacity slicing defined in Sec.~\ref{subsec:supp_candidates}. Only batches already stored at \(i\) in \(\Omega_\tau\), allowed by the TM forwarding rule, and not already reserved in the current half-slot are included. For a TC-carrying action, event \(e\) is eligible only if
	\[
	i\in H_e(\tau),\qquad
	j\notin H_e(\tau),
	\]
	the event is unfinished, and \(i\rightarrow j\) either completes the event or opens a positive downstream frontier according to \(G_e(i,j)\). Eligible TC items are ordered and capacity-sliced by the rules in Sec.~\ref{subsec:supp_candidates}. The resulting planned-item list is frozen on the Stage~1 action and immediately excluded from later duplicate reservation, but its physical service effect is delayed until the half-slot boundary.
	
	For a Stage~1 anchor--anchor or non-anchor--non-anchor contact, neither direction has an inherent ground-access role. Its first-half direction is selected once at the beginning of slot \(t\) from the state available at boundary \(2t\). Let
	\(\Gamma_{i\rightarrow j}^{\TC}(2t)\)
	be the set of unfinished events for which \(i\) is a holder, \(j\) is not a holder, and \(G_e(i,j)>0\) at boundary \(2t\). The directional score is
	\begin{equation}
		\Phi_{i\rightarrow j}^{\TC}(2t)
		=
		\sum_{e\in\Gamma_{i\rightarrow j}^{\TC}(2t)}
		A_e P_e^{\TC}
		M_e^{\TC}(2t)
		G_e(i,j),
		\label{eq:supp_stage1_direction_score}
	\end{equation}
	where \(P_e^{\TC}\) is the destination-type priority, \(M_e^{\TC}(2t)\) is the slot-age multiplier evaluated at the slot-start boundary, and \(G_e(i,j)\) is defined in (\ref{eq:supp_tc_gain}). If
	\(\Phi_{i\rightarrow j}^{\TC}(2t)>
	\Phi_{j\rightarrow i}^{\TC}(2t)\),
	\(i\rightarrow j\) is scheduled in \((t,0)\); the reverse direction is scheduled in \((t,1)\). The reverse ordering is used when the reverse score is larger. Ties are broken deterministically by letting the smaller-index node transmit first.
	
	The orientation is fixed at the slot start, but carried-service items are evaluated separately at the beginning of each half-slot. Therefore, information delivered by the first-half Stage~1 action becomes available after boundary \(2t+1\) and may be eligible for the reverse Stage~1 action or a Stage~2 action in the second half-slot, provided that the corresponding routing and reservation rules are satisfied. Across later half-slots, Stage~1 same-role actions and Stage~2 cross-role TC layers may concatenate into a multi-hop TC path.

	\subsection{Layer Urgency Computation}
	\label{subsec:supp_urgency}
	
	Layer urgency determines the execution order of Stage~2 layers in half-slot \(\tau\); it does not directly select links. Candidate selection is determined later by feasibility filtering, marginal weights, and beam-copy matching. All urgency values are recomputed from the current physical snapshot \(\Omega_\tau\). Navigation layers are active only for \(h=0\); they are omitted from the execution order when \(h=1\).
	
	\subsubsection{Navigation urgency}
	
	Let \(D_u\) be the number of distinct GNSS satellites required by navigation user \(u\) in one superframe. The remaining deficit is
	\begin{equation}
		\Delta_u^{\NAV}(\tau)
		=
		\max\{0,D_u-|Q_u(\tau)|\}.
		\label{eq:supp_nav_deficit}
	\end{equation}
	Define the total remaining navigation demand as
	\begin{equation}
		U^{\NAV}(\tau)
		=
		\sum_{u\in\Uset^{\mathrm{N}}}
		\Delta_u^{\NAV}(\tau).
		\label{eq:supp_nav_total_deficit}
	\end{equation}
	The implementation uses the same layer-ordering urgency for the short- and long-range navigation layers:
	\begin{equation}
		u_{\ell_{\NAV}^{\mathrm{NS}}}(\tau)
		=
		u_{\ell_{\NAV}^{\mathrm{NL}}}(\tau)
		=
		\rho_{\NAV}U^{\NAV}(\tau),
		\label{eq:supp_nav_urgency}
	\end{equation}
	where
	\(\rho_{\NAV}
	=
	\rho_{\ell_{\NAV}^{\mathrm{NS}}}
	=
	\rho_{\ell_{\NAV}^{\mathrm{NL}}}\).
	The user-specific priority \(P_u^{\NAV}\) is not part of layer urgency; it enters the navigation candidate weight in Sec.~\ref{subsec:supp_candidates}.
	
	\subsubsection{Telemetry urgency}
	
	For TM batch \(b\), let \(A_b\) be its current remaining amount, \(P_b^{\TM}\) the priority inherited from its source type, and \(g_b\) its generation slot. Stage~2 executes every half-slot, but TM aging advances in slot units. At half-slot \(\tau=2t+h\),
	\begin{equation}
		a_b(\tau)
		=
		\left\lfloor\frac{\tau}{2}\right\rfloor-g_b.
		\label{eq:supp_tm_slot_age}
	\end{equation}
	Thus, the two half-slots of slot \(t\) use the same batch age \(t-g_b\). The TM age multiplier is
	\begin{equation}
		M_b^{\TM}(\tau)
		=
		\min\left\{
		\overline{M}^{\TM},
		\widetilde{M}_b^{\TM}(\tau)
		\right\},
		\label{eq:supp_tm_age_multiplier}
	\end{equation}
	where
	\begin{equation}
		\begin{aligned}
			\widetilde{M}_b^{\TM}(\tau)
			=
			\exp\!\Bigl(
			&\kappa_1^{\TM}
			\min\{a_b(\tau),d^{\TM}\}\\
			&+\kappa_2^{\TM}
			[a_b(\tau)-d^{\TM}]^+
			\Bigr),
		\end{aligned}
		\label{eq:supp_tm_raw_multiplier}
	\end{equation}
	and \([x]^+=\max\{x,0\}\). Here,
	\(d^{\TM}\) is the slot-age threshold,
	\(\kappa_1^{\TM}\) and
	\(\kappa_2^{\TM}\) are the pre- and post-threshold growth rates, and
	\(\overline{M}^{\TM}\) is the multiplier cap.
	
	The TM-user-access urgency is
	\begin{equation}
		u_{\ell_{\TM}^{\mathrm{U2S}}}(\tau)
		=
		\rho_{\ell_{\TM}^{\mathrm{U2S}}}
		\sum_{u\in\Uset^{\mathrm{M}}}
		B_u^{\TM}(\tau).
		\label{eq:supp_tm_access_urgency}
	\end{equation}
	For any satellite subset
	\(\mathcal{X}\subseteq\Sset\),
	define its weighted TM pressure as
	\begin{equation}
		\Psi_{\TM}(\mathcal{X},\tau)
		=
		\sum_{s\in\mathcal{X}}
		\sum_{b\in\mathcal{B}^{\TM}_s(\tau)}
		A_bP_b^{\TM}M_b^{\TM}(\tau).
		\label{eq:supp_tm_pressure}
	\end{equation}
	The two groundward transfer urgencies are then
	\begin{align}
		u_{\ell_{\TM}^{\mathrm{N2A}}}(\tau)
		&=
		\rho_{\ell_{\TM}^{\mathrm{N2A}}}
		\Psi_{\TM}(\Nset_f,\tau),
		\label{eq:supp_tm_n2a_urgency}\\
		u_{\ell_{\TM}^{\mathrm{A2G}}}(\tau)
		&=
		\rho_{\ell_{\TM}^{\mathrm{A2G}}}
		\Psi_{\TM}(\Aset_f,\tau).
		\label{eq:supp_tm_a2g_urgency}
	\end{align}
	
	\subsubsection{Telecommand urgency}
	
	Let \(P_e^{\TC}\) denote the destination-type priority of event \(e\), determined by whether \(d(e)\) is a GNSS satellite or an external TC user. Let \(P_{\mathrm{U}}^{\TC}\) be the global user-target priority coefficient used in TC layer urgency, and let \(P_u^{\TC,\mathrm{req}}\) be the request priority used only in TC-user delivery candidate weights.
	
	TC aging also advances in slot units. For event \(e\),
	\begin{equation}
		a_e(\tau)
		=
		\left\lfloor\frac{\tau}{2}\right\rfloor-g_e.
		\label{eq:supp_tc_slot_age}
	\end{equation}
	Let \(d_e^{\TC}\) be the slot-age threshold selected according to the destination type. The TC age multiplier is
	\begin{equation}
		M_e^{\TC}(\tau)
		=
		\min\left\{
		\overline{M}^{\TC},
		\widetilde{M}_e^{\TC}(\tau)
		\right\},
		\label{eq:supp_tc_age_multiplier}
	\end{equation}
	with
	\begin{equation}
		\begin{aligned}
			\widetilde{M}_e^{\TC}(\tau)
			&=
			\exp\!\left(
			\kappa_1^{\TC}a_e(\tau)/d_e^{\TC}
			\right),
			&& a_e(\tau)\le d_e^{\TC},
			\\
			&=
			\exp\!\left(
			\kappa_1^{\TC}
			+
			\kappa_2^{\TC}
			[a_e(\tau)-d_e^{\TC}]
			\right),
			&& a_e(\tau)>d_e^{\TC}.
		\end{aligned}
		\label{eq:supp_tc_raw_multiplier}
	\end{equation}
	
	For a possible copy \(i\rightarrow j\), let \(F_e(j,\tau)\) denote the number of currently visible, role-feasible downstream opportunities that would become available after \(j\) joins the holder set, excluding nodes already in \(H_e(\tau)\). A direct visible delivery opportunity to the target is counted as a downstream opportunity when permitted by the TC forwarding rules. The frontier gain is
	\begin{equation}
		G_e^{\mathrm{front}}(j,\tau)
		=
		1+
		\gamma_{\mathrm{front}}
		\log\!\left(1+F_e(j,\tau)\right).
		\label{eq:supp_tc_frontier_gain}
	\end{equation}
	The marginal TC gain is
	\begin{equation}
		G_e(i,j;\tau)
		=
		\begin{cases}
			\gamma_{\mathrm{done}},
			& j=d(e),\\
			G_e^{\mathrm{front}}(j,\tau),
			& j\ne d(e),\ F_e(j,\tau)>0,\\
			0,
			& \text{otherwise}.
		\end{cases}
		\label{eq:supp_tc_gain}
	\end{equation}
	For readability, later equations may write \(G_e(i,j)\) when the half-slot index is clear.
	
	For
	\[
	\ell\in
	\left\{
	\ell_{\TC}^{\mathrm{G2A}},
	\ell_{\TC}^{\mathrm{A2N}},
	\ell_{\TC}^{\mathrm{D}}
	\right\},
	\]
	let \(\mathcal{P}_{e,\ell}(\tau)\) be the set of feasible positive-value opportunities of event \(e\) in layer \(\ell\). For
	\(\ell_{\TC}^{\mathrm{G2A}}\) and
	\(\ell_{\TC}^{\mathrm{A2N}}\),
	the value of opportunity \(i\rightarrow j\) is
	\[
	A_eP_e^{\TC}
	M_e^{\TC}(\tau)
	G_e(i,j;\tau).
	\]
	For \(\ell_{\TC}^{\mathrm{D}}\), the layer-urgency opportunity value is
	\[
	A_eP_{\mathrm{U}}^{\TC}
	M_e^{\TC}(\tau)
	\gamma_{\mathrm{done}}.
	\]
	Let \(V_{e,\ell}^{(q)}(\tau)\) be the \(q\)-th largest value in
	\(\mathcal{P}_{e,\ell}(\tau)\). The raw TC layer pressure is
	\begin{equation}
		\Xi_\ell^{\TC}(\tau)
		=
		\sum_{e\in\Eset_{\mathrm{open}}^{\TC}(\tau)}
		\sum_{q=1}^{
			\min\{
			K_{\TC},
			|\mathcal{P}_{e,\ell}(\tau)|
			\}}
		\frac{V_{e,\ell}^{(q)}(\tau)}{q}.
		\label{eq:supp_tc_raw_urgency}
	\end{equation}
	The TC layer urgency is
	\begin{equation}
		u_\ell(\tau)
		=
		\rho_\ell
		\min\left\{
		\overline{\Xi}^{\TC},
		\Xi_\ell^{\TC}(\tau)
		\right\},
		\label{eq:supp_tc_urgency}
	\end{equation}
	where \(K_{\TC}\) limits how many alternative opportunities of one event contribute to layer urgency and
	\(\overline{\Xi}^{\TC}\) caps the total pressure of one TC layer.

	\subsection{Candidate Filtering, Item Queues, and Marginal Weights}
	\label{subsec:supp_candidates}
	
	For layer \(\ell\) in half-slot \(\tau\), the scheduler first constructs candidate actions from the endpoint roles and directions defined in the main paper. Candidates are then filtered in the following order: FSA visibility, phase compatibility, physical residual-beam availability, per-layer beam cap, positive marginal value, and duplicate-reservation exclusion. State-dependent values are evaluated from \(\Omega_\tau\); items already reserved by Stage~1, earlier Stage~2 layers, or earlier TM matching rounds in the same half-slot are excluded through the working reservation overlay.
	
	\subsubsection{Generic item queue and capacity slicing}
	
	Transfer candidates may contain multiple eligible TM batches or TC events. Let
	\begin{equation}
		\mathcal{I}
		=
		\bigl(
		(A_1,\nu_1),
		(A_2,\nu_2),
		\ldots
		\bigr),
		\qquad
		\nu_1\ge\nu_2\ge\cdots,
		\label{eq:supp_generic_queue}
	\end{equation}
	be an item queue sorted by non-increasing unit value, where \(A_q\) is the currently eligible amount of item \(q\) and \(\nu_q\) is its unit value. The edge weight used in matching is the total queue pressure
	\begin{equation}
		\operatorname{Tot}(\mathcal{I})
		=
		\sum_q A_q\nu_q.
		\label{eq:supp_total_queue_weight}
	\end{equation}
	If the candidate is selected, the link capacity \(c\) is filled greedily in queue order. The amount planned for item \(q\) is
	\begin{equation}
		\delta_q
		=
		\min\left\{
		A_q,
		\left[
		c-\sum_{p<q}\delta_p
		\right]^+
		\right\},
		\label{eq:supp_capacity_slice}
	\end{equation}
	where \([x]^+=\max\{x,0\}\). Scanning stops when the capacity is exhausted or the queue is empty. The value of the frozen planned-item list is
	\begin{equation}
		\operatorname{Val}(\mathcal{I},c)
		=
		\sum_q\delta_q\nu_q.
		\label{eq:supp_sliced_value}
	\end{equation}
	Thus, \(\operatorname{Tot}(\mathcal{I})\) ranks the complete service pressure behind a candidate, whereas
	\(\operatorname{Val}(\mathcal{I},c)\) describes only the items actually reserved after the candidate is selected.
	
	\subsubsection{Navigation candidates}
	
	For a navigation candidate between user \(u\) and satellite \(s\),
	\begin{equation}
		w_{u,s}^{\NAV}(\tau)
		=
		P_u^{\NAV}
		\omega_{\NAV}
		\Delta_u^{\NAV}(\tau).
		\label{eq:supp_nav_weight}
	\end{equation}
	The candidate is retained only if
	\(s\notin Q_u(\tau)\),
	\(\Delta_u^{\NAV}(\tau)>0\),
	the user and satellite are visible, and the required paired phase pattern is feasible. Navigation candidates are generated only for \(h=0\). For a short-range user, either feasible \(\TR\) or \(\RT\) ordering may be used. For a long-range user, the satellite must use \(\RT\), i.e., user-to-satellite reception in \((t,0)\) followed by satellite-to-user transmission in \((t,1)\). Selection reserves both half-slot actions, both beam consumptions, and the corresponding satellite phase order at once.
	
	\subsubsection{Telemetry candidates}
	
	For TM-user access \(u\rightarrow s\),
	\begin{equation}
		w_{u,s}^{\TM,\mathrm{U2S}}(\tau)
		=
		P_u^{\TM}
		B_u^{\TM}(\tau),
		\label{eq:supp_tm_access_weight}
	\end{equation}
	provided that the user backlog remaining after same-half-slot reservations is positive. If selected, the planned access amount is limited by the user--satellite link capacity and by the unreserved user-side backlog.
	
	For a TM transfer candidate \(i\rightarrow j\), eligibility follows the layer route:
	\(i\in\Nset_f,j\in\Aset_f\) for
	\(\ell_{\TM}^{\mathrm{N2A}}\),
	and
	\(i\in\Aset_f,j\in\Gset\) for
	\(\ell_{\TM}^{\mathrm{A2G}}\).
	The unsorted eligible item set and its ordered queue are
	\begin{align}
		\mathcal{J}_{i,j}^{\TM}(\tau)
		&=
		\left\{
		(A_b,\nu_b)
		\,\middle|\,
		\substack{
			b\text{ is stored at }i\text{ in }\Omega_\tau,\\
			b\text{ is route-eligible and not already reserved}
		}
		\right\},
		\label{eq:supp_tm_candidate_items}\\
		\mathcal{I}_{i,j}^{\TM}(\tau)
		&=
		\operatorname{sort}_{\nu_b\downarrow}
		\mathcal{J}_{i,j}^{\TM}(\tau),
		\label{eq:supp_tm_item_queue}
	\end{align}
	where
	\begin{equation}
		\nu_b
		=
		P_b^{\TM}
		M_b^{\TM}(\tau).
		\label{eq:supp_tm_item_unit_value}
	\end{equation}
	The matching weight is
	\begin{equation}
		w_{i,j}^{\TM}(\tau)
		=
		\operatorname{Tot}
		\left(
		\mathcal{I}_{i,j}^{\TM}(\tau)
		\right).
		\label{eq:supp_tm_transfer_weight}
	\end{equation}
	After selection, the planned TM amounts are obtained by applying the capacity-slicing rule to
	\(\mathcal{I}_{i,j}^{\TM}(\tau)\)
	with link capacity \(c_{i,j}\).
	
	\subsubsection{Telecommand candidates}
	
	For TC replica layers
	\(\ell_{\TC}^{\mathrm{G2A}}\) and
	\(\ell_{\TC}^{\mathrm{A2N}}\),
	endpoint roles are, respectively,
	\(i\in\Gset,j\in\Aset_f\)
	and
	\(i\in\Aset_f,j\in\Nset_f\).
	The unsorted positive-gain event set is
	\begin{equation}
		\begin{aligned}
			\mathcal{J}_{i,j}^{\TC,\mathrm{R}}(\tau)
			=
			\bigl\{
			&(A_e,\mu_e^{\mathrm{R}}(i,j))
			\,\big|\,
			e\in\Eset_{\mathrm{open}}^{\TC}(\tau),\\
			&i\in H_e(\tau),\
			j\notin H_e(\tau),\
			G_e(i,j;\tau)>0,\\
			&(e,j)\text{ not reserved}
			\bigr\}.
		\end{aligned}
		\label{eq:supp_tc_replica_candidate_items}
	\end{equation}
	The queue
	\(\mathcal{I}_{i,j}^{\TC,\mathrm{R}}(\tau)\)
	is obtained by sorting this set in non-increasing order of
	\begin{equation}
		\mu_e^{\mathrm{R}}(i,j)
		=
		P_e^{\TC}
		M_e^{\TC}(\tau)
		G_e(i,j;\tau).
		\label{eq:supp_tc_replica_unit_value}
	\end{equation}
	
	For TC-user delivery
	\(i\rightarrow u\)
	in \(\ell_{\TC}^{\mathrm{D}}\),
	the transmitter must be a satellite holder of an unfinished event whose destination is the visible user \(u\). Define
	\begin{equation}
		\begin{aligned}
			\mathcal{J}_{i,u}^{\TC,\mathrm{D}}(\tau)
			=
			\bigl\{
			&(A_e,\mu_e^{\mathrm{D}}(i,u))
			\,\big|\,
			e\in\Eset_{\mathrm{open}}^{\TC}(\tau),\\
			&d(e)=u,\
			i\in H_e(\tau),\
			\chi_{iu}(f)=1,\\
			&e\text{ not reserved for completion}
			\bigr\},
		\end{aligned}
		\label{eq:supp_tc_delivery_candidate_items}
	\end{equation}
	and sort it in non-increasing order of
	\begin{equation}
		\mu_e^{\mathrm{D}}(i,u)
		=
		P_u^{\TC,\mathrm{req}}
		M_e^{\TC}(\tau)
		\gamma_{\mathrm{done}}.
		\label{eq:supp_tc_delivery_unit_value}
	\end{equation}
	For either a replica queue or a delivery queue, the matching weight is
	\begin{equation}
		w_{i,j}^{\TC}(\tau)
		=
		\operatorname{Tot}
		\left(
		\mathcal{I}_{i,j}^{\TC}(\tau)
		\right),
		\label{eq:supp_tc_weight}
	\end{equation}
	where
	\(\mathcal{I}_{i,j}^{\TC}(\tau)\)
	denotes the corresponding sorted queue. If selected, the planned TC items are obtained by capacity slicing with \(c_{i,j}\).

	\subsection{Beam-Copy Runtime Construction and Iterative Execution}
	\label{subsec:supp_matching}
	
	The main paper defines the beam-copy sets and the maximum-weight matching objective. The runtime implementation expands each physical candidate into graph edges over the beam copies currently exposed by
	\(\widehat{R}_{v,r}^{m}(\tau)\).
	All copy-level edges derived from the same physical candidate inherit the same marginal weight.
	
	For navigation, a candidate \((u,s)\) creates edges from user node \(u\) to the available paired copies of satellite \(s\). For TM-user access, candidate \(u\rightarrow s\) creates edges from user node \(u\) to the available Rx copies of \(s\). For TC-user delivery, candidate \(s\rightarrow u\) creates edges from the available Tx copies of \(s\) to user node \(u\). For the four transfer layers
	\[
	\ell_{\TM}^{\mathrm{N2A}},\quad
	\ell_{\TM}^{\mathrm{A2G}},\quad
	\ell_{\TC}^{\mathrm{G2A}},\quad
	\ell_{\TC}^{\mathrm{A2N}},
	\]
	candidate \(i\rightarrow j\) creates edges from one source-state node representing \(i\) to each available Rx copy of \(j\). Consequently, one physical source selects at most one destination in one matching round, whereas a destination may receive from several sources through different Rx copies.
	
	After solving the maximum-weight matching, matched graph edges are mapped back to physical directed actions. The implementation then verifies each action against the current real phase/resource ledger and the working planned-item reservations. Accepted actions are inserted into the schedule, their beam and phase resources are reserved immediately, and their capacity-sliced planned items are frozen. If several graph edges map to actions whose physical reservations would conflict after earlier accepted edges are applied, only actions that remain feasible are retained.
	
	The TM transfer layers
	\(\ell_{\TM}^{\mathrm{N2A}}\) and
	\(\ell_{\TM}^{\mathrm{A2G}}\)
	may execute at most \(I_{\max}\) matching rounds. All rounds of the same layer execution share the cumulative beam-use cap \(\widehat{R}_{v,r}^{m}(\tau)\). After each round, accepted TM amounts are subtracted from a layer-local shadow queue, the real resource ledger is updated, and the consumed beam resources reduce the remaining layer allowance. The next round reconstructs candidates from the same physical snapshot \(\Omega_\tau\) but with the updated residual resources, remaining layer allowance, and shadow availability. This permits a heavily backlogged source to use additional Tx beams across rounds while preventing the same batch amount from being selected twice or the cumulative layer cap from being exceeded. Newly received TM is never inserted into the shadow queue during the half-slot and therefore cannot be forwarded before the boundary commit.
	
	TC layers execute one matching round per half-slot. Their duplicate effects are suppressed by the working reservation overlay: once event \(e\) is planned to make receiver \(j\) a holder, the pair \((e,j)\) is excluded from later same-half-slot replica candidates; once an event is planned for destination completion, later same-half-slot candidates do not reserve the same event for another completion action.
	
	Navigation layers execute only in the first half-slot of a slot. When a navigation match is accepted, the two opposite half-slot actions, the satellite phase order, and the required beam resources in both halves are reserved together. The user--satellite pair is added to \(Q_u\) only after the second directed action has completed at the corresponding half-slot boundary.

	\subsection{Planned-Item Reservation and Half-Slot Commit}
	\label{subsec:supp_commit}
	
	The scheduling order within a half-slot changes only resource availability and working reservations. Physical service state remains frozen at \(\Omega_\tau\) until all selected actions in half-slot \(\tau\) have been determined. At boundary \(\tau+1\), the planned effects of Stage~1 and Stage~2 actions are committed simultaneously.
	
	\subsubsection{Telemetry commit and delay recording}
	
	When a TM action is reserved, each planned amount is marked unavailable to later layers and later matching rounds in the same half-slot. At the boundary, the reserved amount is removed from its source state. For a user-to-satellite action \(u\rightarrow s\), the transmitted amount is subtracted from \(B_u^{\TM}\) and inserted into \(\mathcal{B}_s^{\TM}(\tau+1)\) while preserving the original batch generation slot and source-type priority. For a satellite-to-satellite action \(i\rightarrow j\), the transmitted amount is subtracted from the corresponding source batch at \(i\) and appended to the same batch identity at \(j\). For a satellite-to-GS action, the transmitted amount is removed from the source batch and recorded as delivered. A batch may therefore be split across actions or boundaries, but every delivered portion retains the generation slot \(g_b\) of the original batch.
	
	If a portion of batch \(b\) reaches a GS at boundary \(b_b^{\mathrm{del}}\), its delivery delay is
	\begin{equation}
		D_b^{\TM}
		=
		b_b^{\mathrm{del}}-2g_b.
		\label{eq:supp_tm_delivery_delay}
	\end{equation}
	When a batch is delivered in several portions, delay statistics are recorded for the delivered amounts at their respective delivery boundaries.
	
	\subsubsection{Telecommand commit, first arrival, and replica pruning}
	
	A planned TC copy \(i\rightarrow j\) does not modify \(H_e(\tau)\) during the half-slot. At boundary \(\tau+1\), all successfully planned receiver nodes are unioned into the holder set:
	\[
	H_e(\tau+1)
	=
	H_e(\tau)
	\cup
	H_e^{\mathrm{new}}(\tau+1),
	\]
	unless the event is completed at that boundary. Multiple simultaneous copies of one event may create multiple new holders, but the original event amount is still counted only once.
	
	Let \(b_e^{\mathrm{done}}\) denote the first boundary at which event \(e\) reaches its destination. If an action in half-slot \(\tau\) delivers the event, its arrival boundary is \(\tau+1\), and
	\begin{equation}
		b_e^{\mathrm{done}}
		=
		\min\left\{
		\tau+1
		\,\middle|\,
		\substack{
			e\text{ reaches }d(e)\\
			\text{through an action in half-slot }\tau
		}
		\right\}.
		\label{eq:supp_tc_first_arrival}
	\end{equation}
	The completion delay is
	\begin{equation}
		D_e^{\TC}
		=
		b_e^{\mathrm{done}}-2g_e.
		\label{eq:supp_tc_completion_delay}
	\end{equation}
	If several replicas reach the destination at the same earliest boundary, the event is completed once and its original amount \(A_e\) is counted once. If no delivery reaches the target before the evaluation boundary, \(b_e^{\mathrm{done}}\) remains undefined and the event is counted as unfinished in that evaluation window.
	
	After completion, all remaining replicas of event \(e\) are removed from active scheduling state. For an unfinished event, redundant holders are pruned after boundary propagation. For holder \(h\), let its valid downstream-neighbor set contain the nodes that are visible from \(h\) and can receive a replica under the TC forwarding rules in the current FSA. If this set is nonempty and every such neighbor already belongs to the updated holder set, the copy at \(h\) is removed because it cannot expand the current reachable frontier.
	
	\subsubsection{Navigation commit}
	
	A navigation service changes the distinct-ranging record only after both required directed actions have completed. For a user--satellite service selected in slot \(t\), the pair is added to \(Q_u\) after the boundary following the second half-slot action. Repeated service with the same user--satellite pair in the same superframe does not increase \(|Q_u|\). The same distinct-counterpart principle is used for GNSS inter-satellite ranging statistics.

	\section{Supplementary Material: Stage~2 Preference Parameters and Tuning Rules}
	\label{sec:supp_preference_parameters}
	
	This section reports the numerical preference parameters used in the evaluation and explains their roles in the runtime calculations of Sec.~\ref{sec:supp_operational_details}. The parameters act at different decision levels and should not be interpreted as interchangeable priority weights. The layer-ordering coefficients \(\rho_\ell\) affect only the relative execution order of Stage~2 layers. The resource-sharing parameters \(\alpha_r\), \(\beta_\ell\), and \(I_{\max}\) control how much residual multibeam capacity a layer can exploit and how many matching rounds the TM transfer layers may execute. Service-value parameters determine navigation candidate weights, TM pressure and batch ordering, and TC event urgency and marginal reachability gain.
	
	Although these parameters are introduced primarily for Stage~2, some service-value terms are intentionally reused by the Stage~1 carried-service mechanism defined in Sec.~\ref{subsec:supp_stage1_actions}. In particular, TM source-type priorities and TM age multipliers are reused when an oriented Stage~1 action carries eligible TM items, while TC event priorities, age multipliers, and completion/frontier gains are reused in TC item selection and in the direction score of same-role Stage~1 satellite contacts. None of these parameters changes the Stage~1 ILP candidate set, constraints, or objective.
	
	\subsection{Dynamic Layer-Ordering Coefficients}
	\label{subsec:supp_order_params}
	
	The urgency functions in Sec.~\ref{subsec:supp_urgency} produce the state-dependent pressure of each Stage~2 layer. The coefficient \(\rho_\ell\) scales that pressure before layers are sorted:
	\[
	u_\ell(\tau)=\rho_\ell\,\widetilde{u}_\ell(\tau),
	\]
	where \(\widetilde{u}_\ell(\tau)\) denotes the unscaled state value defined by the corresponding urgency formula. Therefore, \(\rho_\ell\) affects \emph{inter-layer ordering}; it does not directly change the marginal weights used by maximum-weight matching inside a layer.
	
	Table~\ref{tab:supp_order_params} lists the coefficients used in the evaluation.
	
	\begin{table*}[t]
		\centering
		\caption{Dynamic layer-ordering coefficients used in the evaluation.}
		\label{tab:supp_order_params}
		\footnotesize
		\begin{tabular}{p{0.24\textwidth}p{0.12\textwidth}p{0.52\textwidth}}
			\toprule
			Layer \(\ell\) & \(\rho_\ell\) & Tuning effect \\
			\midrule
			\(\ell_{\NAV}^{\mathrm{NS}}\) &
			10.0 &
			Scales short-range navigation deficit pressure and increases its opportunity to obtain an earlier execution rank. \\
			\(\ell_{\NAV}^{\mathrm{NL}}\) &
			10.0 &
			Scales long-range navigation deficit pressure under the stricter \(\RT\)-compatible service requirement. \\
			\(\ell_{\TM}^{\mathrm{U2S}}\) &
			2.0 &
			Increases the ordering pressure of TM-user access as user-side backlog grows. \\
			\(\ell_{\TM}^{\mathrm{N2A}}\) &
			2.0 &
			Increases the ordering pressure of non-anchor TM transfer toward anchors. \\
			\(\ell_{\TM}^{\mathrm{A2G}}\) &
			2.0 &
			Increases the ordering pressure of anchor-side TM downlink to GSs. \\
			\(\ell_{\TC}^{\mathrm{G2A}}\) &
			3.0 &
			Increases the ordering pressure of TC injection from GSs to visible anchors. \\
			\(\ell_{\TC}^{\mathrm{A2N}}\) &
			3.0 &
			Increases the ordering pressure of TC dissemination from anchors to non-anchors. \\
			\(\ell_{\TC}^{\mathrm{D}}\) &
			3.0 &
			Increases the ordering pressure of direct delivery to visible TC users. \\
			\bottomrule
		\end{tabular}
	\end{table*}
	
	The larger navigation coefficient reflects the finite-superframe ranging objective: an unsatisfied distinct-ranging opportunity disappears when the superframe ends. The TC coefficients are larger than the TM coefficients because TC probes represent delay-sensitive reachability, whereas TM is sustained throughput-oriented traffic whose pressure also accumulates through backlog and age. These coefficients should be tuned as \emph{relative} scales. Excessively increasing one \(\rho_\ell\) can give the corresponding layer repeated early access to residual resources, while an excessively small value can postpone that layer even when useful candidates exist.
	
	The two navigation layers use the same coefficient and, according to (\ref{eq:supp_nav_urgency}), the same total navigation-deficit state value. Their urgency scale is therefore intentionally aligned; their different service outcomes arise from candidate feasibility, especially the stricter \(\RT\) requirement of long-range navigation.
	
	\subsection{Layer Resource Fractions and Matching Rounds}
	\label{subsec:supp_resource_params}
	
	The resource-control parameters instantiate the rank-dependent fraction \(\alpha_r\), layer-dependent fraction \(\beta_\ell\), and TM matching-round limit \(I_{\max}\) used by the main-paper resource cap and Sec.~\ref{subsec:supp_matching}. Their values are listed in Table~\ref{tab:supp_resource_params}.
	
	\begin{table*}[t]
		\centering
		\caption{Resource-sharing and matching-round parameters used in Stage~2.}
		\label{tab:supp_resource_params}
		\footnotesize
		\begin{tabular}{p{0.24\textwidth}p{0.12\textwidth}p{0.52\textwidth}}
			\toprule
			Parameter & Value & Tuning effect \\
			\midrule
			\(\alpha_1\) &
			0.70 &
			Limits the nominal resource share of the first-ranked layer and preserves part of the physical capacity for later layers. \\
			\(\alpha_r,\ r\ge2\) &
			1.00 &
			Applies no additional rank-based cap beyond the real residual-resource ledger. \\
			\(\beta_{\ell_{\NAV}^{\mathrm{NS}}}\) &
			0.70 &
			Limits short-range navigation use of a satellite's Tx/Rx resources. \\
			\(\beta_{\ell_{\NAV}^{\mathrm{NL}}}\) &
			0.70 &
			Limits long-range navigation use of a satellite's Tx/Rx resources. \\
			\(\beta_\ell\) for other layers &
			1.00 &
			Applies no additional layer-specific cap beyond the rank cap and residual resources. \\
			\(I_{\max}\) &
			3 &
			Allows the N2A and A2G TM transfer layers to reuse remaining Tx resources through repeated matching, at additional runtime cost. \\
			\bottomrule
		\end{tabular}
	\end{table*}
	
	With these values, the effective fraction in the main-paper definition
	\(\eta_{\ell_r,r}=\min\{\alpha_r,\beta_{\ell_r}\}\)
	reduces to
	\begin{equation}
		\eta_{\ell_r,r}
		=
		\begin{cases}
			0.70,
			&
			r=1
			\ \text{or}\
			\ell_r\in
			\{\ell_{\NAV}^{\mathrm{NS}},
			\ell_{\NAV}^{\mathrm{NL}}\},
			\\[1mm]
			1.00,
			&
			r\ge2
			\ \text{and}\
			\ell_r\notin
			\{\ell_{\NAV}^{\mathrm{NS}},
			\ell_{\NAV}^{\mathrm{NL}}\}.
		\end{cases}
		\label{eq:supp_numeric_effective_fraction}
	\end{equation}
	The one-beam minimum in the resource-cap rule remains active whenever the effective fraction is positive and at least one real residual beam is available. Hence, the 70\% setting is a nominal absolute-cap fraction of physical beam capacity, not a strict fractional bound for one- or two-beam terminals.
	
	The role of \(\alpha_r\) is cross-layer coexistence: a smaller \(\alpha_1\) preserves more beams for later layers, whereas a larger value gives the most urgent layer a stronger immediate response. The role of \(\beta_\ell\) is functional protection independent of rank. In the evaluated setting, only navigation uses an additional functional cap because one navigation service reserves both opposite half-slot actions and consumes paired Tx/Rx resources of the serving satellite.
	
	The iterative limit applies only to
	\(\ell_{\TM}^{\mathrm{N2A}}\)
	and
	\(\ell_{\TM}^{\mathrm{A2G}}\).
	All other Stage~2 layers execute one matching round per eligible half-slot. Increasing \(I_{\max}\) can improve Tx-beam utilization at heavily backlogged TM sources, but it increases candidate reconstruction and matching cost and may leave fewer residual resources to later layers.
	
	\subsection{Navigation and Telemetry Weight Parameters}
	\label{subsec:supp_nav_tm_params}
	
	The parameters in this subsection affect navigation candidate weights and TM backlog valuation. They do not determine Stage~2 layer order directly; layer order is controlled by the urgency functions and \(\rho_\ell\).
	
	Table~\ref{tab:supp_nav_tm_params} lists the values used in the evaluation.
	
	\begin{table*}[t]
		\centering
		\caption{Navigation and TM weighting parameters used in the evaluation.}
		\label{tab:supp_nav_tm_params}
		\footnotesize
		\begin{tabular}{p{0.24\textwidth}p{0.12\textwidth}p{0.52\textwidth}}
			\toprule
			Parameter & Value & Tuning effect \\
			\midrule
			\(\omega_{\NAV}\) &
			5.0 &
			Sets the common scale of navigation candidate weights; relative NAV-user preference is determined by request priority and remaining distinct-ranging deficit. \\
			\(P_u^{\NAV}\) &
			1.0 &
			Uniform default NAV-user request priority; heterogeneous values can differentiate user importance within a navigation layer. \\
			\(P_u^{\TM}\) &
			1.0 &
			Uniform default TM-user request priority in U2S access matching. \\
			\(P_{\mathrm{sat}}^{\TM}\) &
			3.0 &
			Source-type priority inherited by TM generated by GNSS satellites. \\
			\(P_{\mathrm{usr}}^{\TM}\) &
			1.0 &
			Source-type priority inherited by TM generated by external TM users after admission to the constellation. \\
			\(d^{\TM}\) &
			12 &
			Slot-age threshold separating the slow- and fast-aging regions. \\
			\(\kappa_1^{\TM}\) &
			0.02 &
			Controls TM value growth before the age threshold. \\
			\(\kappa_2^{\TM}\) &
			0.12 &
			Controls faster growth after the threshold and therefore long-tail delay suppression. \\
			\(\overline{M}^{\TM}\) &
			1000.0 &
			Caps the influence of very old TM batches on pressure and transfer weights. \\
			\bottomrule
		\end{tabular}
	\end{table*}
	
	For a batch \(b\), the source-type priority used in
	(\ref{eq:supp_tm_pressure}) and
	(\ref{eq:supp_tm_item_unit_value}) is
	\begin{equation}
		P_b^{\TM}
		=
		\begin{cases}
			P_{\mathrm{sat}}^{\TM},
			& b\text{ originates at a GNSS satellite},\\
			P_{\mathrm{usr}}^{\TM},
			& b\text{ originates at an external TM user}.
		\end{cases}
		\label{eq:supp_tm_source_priority}
	\end{equation}
	The priority remains attached to the batch as it moves between satellites. Thus, satellite-generated TM and admitted user TM compete through their original source type rather than the identity of the current relay.
	
	The NAV weight scale \(\omega_{\NAV}\) is common to all navigation candidates. The within-layer preference between users is therefore created by \(P_u^{\NAV}\Delta_u^{\NAV}(\tau)\), while inter-layer access order is governed separately by navigation urgency and \(\rho_{\NAV}\). In the evaluation, \(P_u^{\NAV}=1\) for all navigation users, so remaining distinct-ranging deficit is the only user-dependent weight term.
	
	For TM-user access, the U2S candidate weight in
	(\ref{eq:supp_tm_access_weight})
	depends on user-side backlog and \(P_u^{\TM}\); the TM age multiplier is not applied before the data enter the constellation. After admission, the source-type priority and age multiplier affect N2A/A2G urgency, transfer-edge weights, and the value ordering of TM items carried by eligible Stage~1 actions. Raising
	\(P_{\mathrm{sat}}^{\TM}\)
	protects GNSS self-telemetry under congestion, whereas raising
	\(P_{\mathrm{usr}}^{\TM}\)
	gives admitted external-user data stronger competition for satellite-side transfer capacity.
	
	The aging parameters control starvation protection and tail delay. Decreasing \(d^{\TM}\) moves batches into the faster-growth region earlier; increasing \(\kappa_2^{\TM}\) more aggressively raises the value of overdue batches. Excessively aggressive aging can reduce throughput-oriented efficiency by allowing a small amount of very old traffic to dominate transfer decisions, while overly weak aging can leave long-waiting batches persistently behind large fresh backlogs.
	
	\subsection{Telecommand Event-Weighting Parameters}
	\label{subsec:supp_tc_params}
	
	TC parameters act at several related locations: TC layer urgency, TC replica and delivery candidate weights, Stage~1 TC carried-item ordering, and the direction score of Stage~1 same-role satellite contacts. Table~\ref{tab:supp_tc_params} lists the values used in the evaluation.
	
	\begin{table*}[t]
		\centering
		\caption{TC event-weighting parameters used in the evaluation.}
		\label{tab:supp_tc_params}
		\footnotesize
		\begin{tabular}{p{0.24\textwidth}p{0.12\textwidth}p{0.52\textwidth}}
			\toprule
			Parameter & Value & Tuning effect \\
			\midrule
			\(P_{\mathrm{sat}}^{\TC}\) &
			6.0 &
			Destination-type priority of events targeting GNSS satellites. \\
			\(P_{\mathrm{U}}^{\TC}\) &
			4.0 &
			Destination-type priority of user-target events in TC propagation and layer urgency. \\
			\(P_u^{\TC,\mathrm{req}}\) &
			1.0 &
			Request priority used specifically in satellite-to-user TC delivery candidate weights. \\
			\(d_{\mathrm{sat}}^{\TC}\) &
			4 &
			Slot-age threshold for satellite-target events. \\
			\(d_{\mathrm{U}}^{\TC}\) &
			7 &
			Slot-age threshold for user-target events. \\
			\(\kappa_1^{\TC}\) &
			0.10 &
			Controls pre-threshold TC age growth. \\
			\(\kappa_2^{\TC}\) &
			0.35 &
			Controls strong post-threshold prioritization of delayed events. \\
			\(\overline{M}^{\TC}\) &
			\(10^6\) &
			Caps the age multiplier used by TC urgency and candidate values. \\
			\(\overline{\Xi}^{\TC}\) &
			5000.0 &
			Caps the raw pressure of one TC layer before multiplication by \(\rho_\ell\). \\
			\(\gamma_{\mathrm{done}}\) &
			100.0 &
			Gives a large bonus to actions that directly complete an event. \\
			\(\gamma_{\mathrm{front}}\) &
			0.25 &
			Controls the additional value of creating a holder with more downstream opportunities. \\
			\(K_{\TC}\) &
			1 &
			Limits how many alternative opportunities of one event contribute to TC layer urgency. \\
			\bottomrule
		\end{tabular}
	\end{table*}
	
	For event \(e\), the destination-type priority and age threshold used in the propagation terms are
	\begin{align}
		P_e^{\TC}
		&=
		\begin{cases}
			P_{\mathrm{sat}}^{\TC},
			& d(e)\in\Sset,\\
			P_{\mathrm{U}}^{\TC},
			& d(e)\in\Uset^{\mathrm{C}},
		\end{cases}
		\label{eq:supp_tc_destination_priority}\\
		d_e^{\TC}
		&=
		\begin{cases}
			d_{\mathrm{sat}}^{\TC},
			& d(e)\in\Sset,\\
			d_{\mathrm{U}}^{\TC},
			& d(e)\in\Uset^{\mathrm{C}}.
		\end{cases}
		\label{eq:supp_tc_destination_threshold}
	\end{align}
	These destination-type terms are distinct from
	\(P_u^{\TC,\mathrm{req}}\), which is used in the final satellite-to-user delivery weight in
	(\ref{eq:supp_tc_delivery_unit_value}).
	
	The age parameters affect both TC layer urgency and candidate values through
	\(M_e^{\TC}(\tau)\).
	The cap
	\(\overline{\Xi}^{\TC}\)
	acts only on the aggregate layer pressure in
	(\ref{eq:supp_tc_urgency}); it does not cap individual candidate weights. Likewise,
	\(K_{\TC}\)
	controls how many top opportunities of one event contribute to layer urgency, but it does not limit how many feasible TC candidates can be constructed.
	
	The completion and frontier coefficients control the balance between immediate delivery and proactive reachability expansion. A larger
	\(\gamma_{\mathrm{done}}\)
	more strongly favors actions that reach the target directly. A larger
	\(\gamma_{\mathrm{front}}\)
	increases the value of copying an event to a node that opens additional downstream opportunities. Because the same gain function \(G_e(i,j;\tau)\) is reused in Stage~2 TC urgency, TC replica candidate weights, Stage~1 TC item valuation, and same-role Stage~1 orientation scores, these two coefficients should be tuned consistently rather than independently for each scheduling stage.
	
	The choice \(K_{\TC}=1\) means that, for layer ordering, each unfinished event contributes only its best currently feasible opportunity in a TC layer. This event-level top-one rule prevents an event with many simultaneously visible propagation options from creating artificially large layer urgency merely because it has many alternative outgoing links.
	
	\subsection{Recommended Tuning Order}
	\label{subsec:supp_tuning_order}
	
	The parameters operate at different levels of the scheduler and are best tuned in that order. First, \(\alpha_r\), \(\beta_\ell\), and \(I_{\max}\) should be chosen to match the desired cross-layer sharing and computational budget. Second, the layer-ordering coefficients \(\rho_\ell\) should be adjusted to set the relative response speed of NAV, TM, and TC layers under comparable state pressure. Third, request and source-type priorities should be tuned to express service-class protection and differentiated user importance. Finally, aging thresholds and growth rates should be adjusted to control tail delay, followed by \(\gamma_{\mathrm{done}}\) and \(\gamma_{\mathrm{front}}\) to balance immediate TC completion against proactive frontier expansion.
	
	This ordering separates four different effects:
	\[
	\begin{aligned}
		\text{resource exposure}
		&\rightarrow
		\text{inter-layer order}\\
		&\rightarrow
		\text{within-layer/service-item preference}\\
		&\rightarrow
		\text{delay and reachability shaping}.
	\end{aligned}
	\]
	Changing several levels simultaneously can obscure which mechanism causes a performance shift. The values reported in Tables~\ref{tab:supp_order_params}--\ref{tab:supp_tc_params} are the parameter set used for the evaluation.

\end{document}